\documentclass[%
reprint,
 amsmath,amssymb,
 aps,
]{revtex4-2}

\usepackage{graphicx}
\usepackage{dcolumn}
\usepackage{bm}
\usepackage{hyperref}

\newcommand{\grad}{{\bf \nabla}}
\newcommand{\xv}{{\bf x}}
\newcommand{\nv}{{\bf n}}
\newcommand{\Nv}{{\bf N}}
\newcommand{\ev}{{\bf \hat{e}}}
\newcommand{\mv}{{\bf m}}

\begin{document}

\preprint{APS/123-QED}

\title{Dispersed, condensed and self-limiting states of geometrically frustrated assembly}

\author{Nicholas W. Hackney}
\affiliation{Department of Physics, University of Massachusetts Amherst}
\author{Christopher Amey}
\affiliation{Department of Physics, Brandeis University}
\author{Gregory M. Grason}
\affiliation{Department of Polymer Science and Engineering, University of Massachusetts Amherst}

\date{\today}

\begin{abstract}
In self-assembling systems, geometric frustration leads to complex states characterized by internal gradients of shape misfit.  Frustrated assemblies have drawn recent interest due to the unique possibility that their thermodynamics can sense and select the finite size of assembly at length scales much larger than constituent building blocks or their interactions.  At present, self-limitation is chiefly understood to derive from zero-temperature considerations, specifically the competition between cohesion and scale-dependent elastic costs of frustration. While effects of entropy and finite temperature fluctuations are necessarily significant for self-assembling systems, their impact on the self-limiting states of frustrated assemblies is not known.  We introduce a generic, minimal model of frustrated assembly, and establish its finite-temperature and concentration dependent thermodynamics by way of simulation and continuum theory.  The phase diagram is marked by three distinct states of translation order:  a dispersed vapor; a defect-riddled condensate; and the self-limiting aggregate state.  We show that, at finite temperature, the self-limiting state is stable at intermediate frustration. Further, in contrast to the prevailing picture, its thermodynamic boundaries with the macroscopic disperse and bulk states are temperature controlled, pointing to the essential importance of translational and conformational entropy in their formation.
\end{abstract}

\maketitle


\begin{figure*}[ht]%
\includegraphics[width=\textwidth]{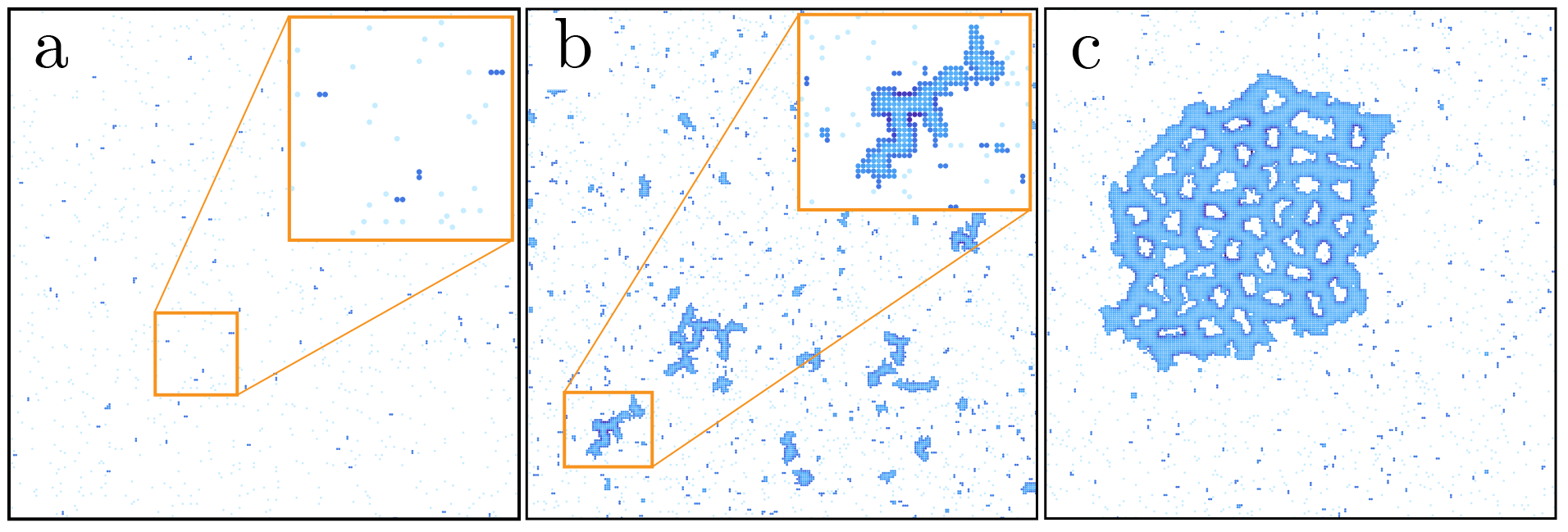}
\caption{\textbf{States of Frustrated Assembly } Monte Carlo simulation snapshots depicting the three hallmark states of equilibrium frustrated assembly: \textbf{a} dispersed vapor free monomers at dilute and highly-frustrated conditions \textbf{b} finite-width aggregates in saturated systems a intermediate frustration and \textbf{c} bulk phase separation (vortex sponge condensate) at weak frustration. Subunits are colored by the strain energy as in (Fig. \ref{fig: model}).} \label{fig: states}
\end{figure*}

\section{Introduction}

Geometric frustration is a common feature of many condensed matter systems, from low-temperature magnetism to liquid crystals, widely associated with the obstruction of defect-free bulk order in the ground state~\cite{sadoc}.  In self-assembling soft matter systems, geometric frustration leads to distinct scale-dependent behaviors absent from bulk systems~\cite{grason2016perspective}, emerging from additional degrees of freedom associated with reconfigurable boundaries of a finite-size assembly~\cite{witten}. In these systems, the costs of local shape misfit propagate to large length scales~\cite{grason2017misfits, meiri2021cumulative} and can compete with cohesive drive to increase size, leading to what is arguably the most salient behavior: {\it self-limiting aggregation} (SLA), where equilibrium assembly sizes are finite yet much larger than subunits or their interactions~\cite{HaganGrason}.  The ability of frustrated assemblies to limit their size on the supra-particle scale distinguishes them from more canonical forms of assembly, like micellar aggregates that are limited in thickness by the size of their amphiphilic constituents~\cite{israelachvili_intermolecular_1992}.  This paradigm has been used to rationalize structural observations in a diverse range of physical systems, including two-dimensional crystallization on curved surfaces~\cite{schneider2005shapes, meng, Das2022}, assemblies of tetrahedral nanoparticles~\cite{serafin}, ribbons of chiral amphiphiles~\cite{GhafBruin, Sharon}, twisted protein fibers~\cite{aggeli2001hierarchical, Turner2003, hall2016morphology, Grason2020}, spherical protein shells~\cite{Mendoza, li2018large} and geometrically incompatible polygonal particles~\cite{witten, spivack2022stress, tyukodi2022thermodynamic}.  Beyond existing systems, the unique possibility of engineering the finite size of a self-assembling system via frustrated building blocks poses new opportunities for ``programming'' self-assembly by design~\cite{HaganGrason}, for example exploiting recent advances in DNA nanotechnology for intentionally shape misfitting colloids~\cite{tyukodi2022thermodynamic, Berengut2020}.

To date, understanding of self-limiting aggregation in frustrated assembly relies almost exclusively on zero-temperature, continuum theories~\cite{grason2016perspective} that pit the competing effects of intra-assembly elastic gradients of strain (i.e. shape misfit) against the cohesive cost of boundary formation on the ground state morphology. Notably this picture leaves out the entropic considerations underpinning equilibrium assembly at finite temperature and concentration, effects which seemingly pose a basic paradox for the putative existence of SLA. On one hand, assembly thermodynamics requires chemical equilibrium between aggregated and free subunits, implying that subunits join or leave aggregates at the $k_B T$-energy scale. On the other hand, self-limitation depends on the propagation of elastic gradients throughout the intra-assembly order which must not be melted away by finite temperature fluctuations. Moreover, the existence of SLA of non-trivial size requires that the super-extensive elastic cost of frustration is not superseded by the incorporation of localized defects~\cite{jellium}, which are necessary features of bulk frustrated order~\cite{kleman}. Taken together, these considerations raise basic questions about whether and under what conditions, this equilibrium self-limitation occurs in frustrated assembly.  For example, a recent simulation study of a discrete particle model of frustrated hyperbolic tubules showed that the finite-size free-energy minimum survives for at least specific non-zero temperature values~\cite{tyukodi2022thermodynamic}, yet it remains unknown what determines the range of thermal stability of this self-limiting state.  

\begin{figure*}[ht]%
\includegraphics[width=\textwidth]{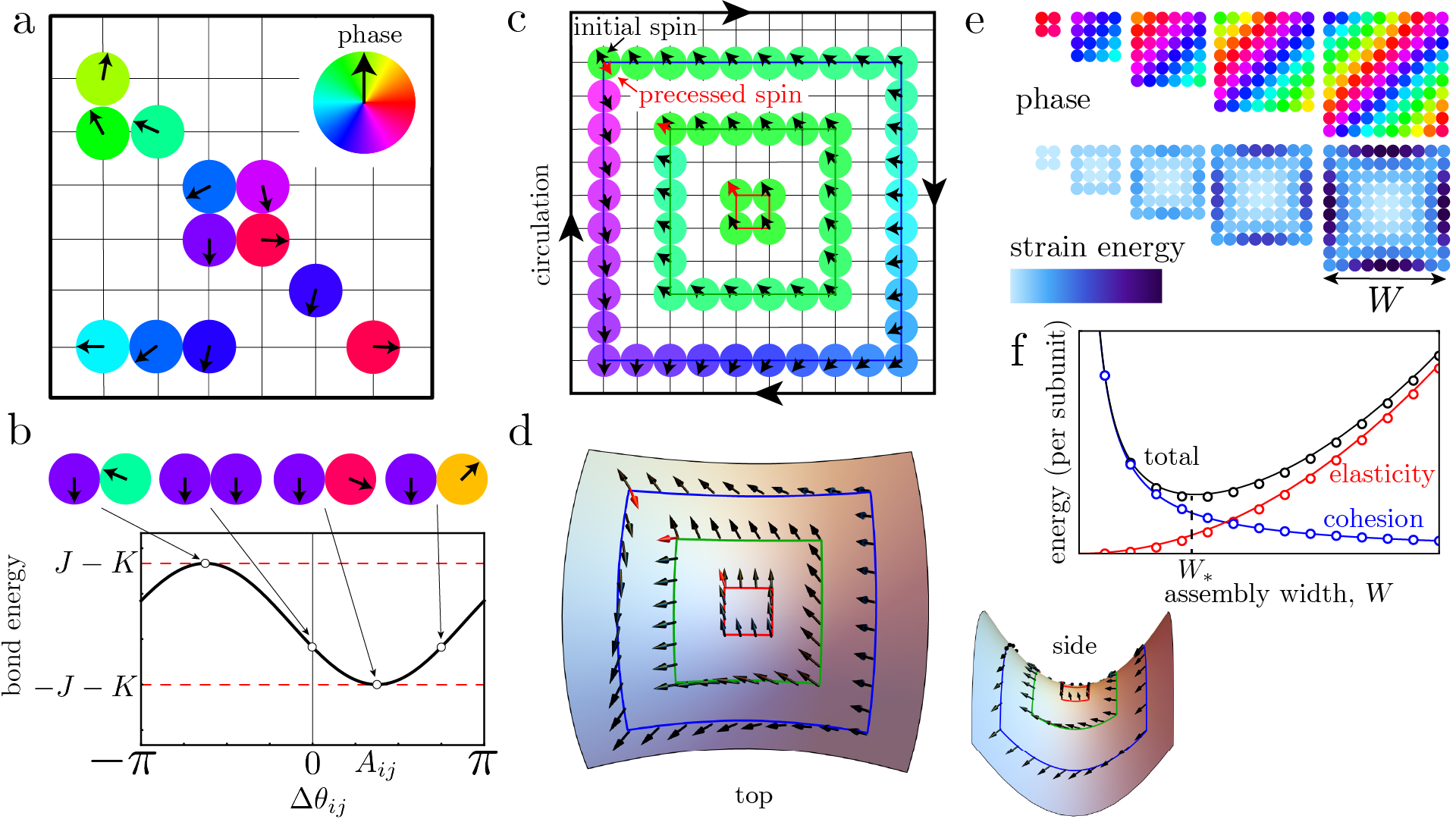}
\caption{\textbf{Lattice model of frustrated assembly.} \textbf{a} Schematic illustration of lattice degrees of freedom, with orientation (phase) shown both as vector and color wheel.  \textbf{b} Schematic of interactions between neighbor subunits, dependent on their relative phase difference $\Delta \theta_{ij}=\theta_{i} -\theta_{j}$ which favors a preferred phase misfit $A_{ij}$ (defined on lattice bonds). \textbf{c} Schematic illustration of size-dependent effects of the local frustration defined in eq. \ref{eq: GaugeSum} for closed loops of occupied bonds.  Spins follow optimal phase misfit along clockwise loops starting and ending at the upper left, resulting mismatch between initial (black) and optimally precessed  (red) spin which grows with loops area.  \textbf{d} Illustration of analogous frustration of vectorial order on a hyperbolic surface, where optimal precession is defined by parallel transport.  \textbf{e} Ground states of square clusters of bound subunits of increasing size $W$ showing both phase gradients (top) as well as gradients in phase strain energy (bottom). Note that the diagonally striped pattern in the phase gradients is a consequence of the particular choice of gauge, while the gradient patterns of the elastic phase strain are gauge invariant (i.e. under transforms which maintain $\grad \times {\bf A}$). \textbf{f} Schematic illustration of the competing effects of cohesion (edge energy) and elasticity (frustration) on the energy density of clusters and the selection of an optimal finite size $W_*$. Solid lines correspond to theoretical results for the energy density of frustrated square-shaped assemblies and the open markers correspond to numerically measured energy density for square aggregates (Fig. \ref{fig: model}e).
} \label{fig: model}
\end{figure*}

Here we study a minimal lattice model of geometrically frustrated assembly and show that its finite-temperature behavior is marked by three distinct states of translational order shown in Fig.~\ref{fig: states}: (a) a dispersed state dominated by free subunits; (b) a self-limiting state, where dominant aggregates are characterized by multi-unit finite-widths but exhibit variable lengths and conformations; and (c) phase separation into a sponge-like bulk condensate which is unlimited in size but perforated by quasi-regular arrays of topologically-charged voids (vortices). Equilibrium SLA is delimited by dispersion at large frustration and low concentration and by a thermodynamic transition to a defective state of bulk aggregation at low frustration, with both of these transitions exhibiting a strong temperature dependence.

\section{Lattice model of frustrated assembly} We introduce a minimal 2D lattice model for self-assembly in frustrated systems, shown schematically in Fig.~\ref{fig: model}a. This model contains two essential ingredients:  translational degrees of freedom for cohesive subunits and energy costs of ``shape misfit'' favored by local interactions (see Appendix~\ref{sim_appendix} for full details).  We consider a square lattice with a fixed fraction ($\Phi$) of sites occupied by assembling units, where $\eta_i = 0,1$ is the occupancy at site $i$.  As a minimal description of continuously deformable local shape, each occupied site also carries a rotational degree of freedom, or phase, $\theta_i \in [0,2 \pi]$.  The system is described by the lattice Hamiltonian
\begin{equation}\label{eq: Hamiltonian}
    H=-J\sum_{\langle ij\rangle}\cos\big(\Delta \theta_{ij}-A_{ij}\big)\eta_i\eta_j -K\sum_{\langle ij\rangle}\eta_i\eta_j,
\end{equation}
where $\langle ij\rangle$ denotes nearest neighbor lattice sites, $\Delta \theta_{ij}=\theta_i-\theta_j$ is the difference in phase between  $i$ and $j$.  The second term encodes an Ising-like, non-specific binding energy $K$ between neighbors while the first term describes an orientationally-dependent neighbor cohesion.  Frustration is introduced through a preferred misalignment between neighbors, $\Delta \theta_{ij}=A_{ij}$, encoded in a gauge field defined on lattice bonds, $A_{ij}$, leading to an orientationally-dependent cohesion shown schematically in Fig.~\ref{fig: model}b.  For fully occupied XY models ($\Phi = 1$)~\cite{teitelprl1983, teitelprb1983}, it is known that when preferred misfits $A_{ij}$ do not sum to integer multiples of $2\pi$ around elementary plaquettes, phase strain is unavoidable.  Frustration strength $f$ is defined by the (orientated) plaquette sum,
\begin{equation}\label{eq: GaugeSum}
    \sum_{\circlearrowright}A_{ij}=2\pi f .
\end{equation}
Here we consider the case of fixed and uniform frustration~\cite{teitelprb1983, tarjus2005frustration}, which can be realized by $A_{ij} = \int_{{\bf x}_i}^{{\bf x}_j} {\rm d}{\bf x} \cdot {\bf A}({\bf x})$, where ${\bf A}({\bf x})$ is a 2D vector field $\nabla_\perp \times {\bf A} = 2 \pi f$.

The scale dependent nature of frustration in the model can be illustrated by considering the winding of spins in closed loops, as shown in Fig.~\ref{fig: model}c.  Precessing $\theta_i$ in a clockwise loop according to the preferred rotation $A_{ij}$ leads to a net rotation of spin relative to the starting point, an accumulated spin misfit that grows with loop area.  In fully occupied, uniformly frustrated XY models, frustration corresponds to the interval $-1/2<f<1/2$, and the limit of $f \ll 1$ results in Abrikosov-like ground states, punctuated by arrays of vortex defects at densities proportional to $f$, implying a characteristic vortex spacing $\ell_{\rm v} \sim f^{-1/2}$ that grows large in the limit of small frustration ~\cite{franz1995vortex, tarjus2005frustration, alba2008uniformly}.  

In dilute regimes of this model ($\Phi \ll 1$), a distinct class of finite-domain, defect-free ground states is possible. These states mitigate the accumulation of elastic costs due the presence of free boundaries that screen the far-field effects of frustration (see e.g. Fig.~\ref{fig: model}e).  The energy of such a domain ${\cal D}$ of fully occupied sites that is much larger than the lattice spacing is well described by the continuum energy,
\begin{equation}
\label{eq: continuum}
E[{\cal D}]= \frac{J}{2} \int_{ {\cal D} } {\rm d}^2{\bf x} ~ \big\lvert \nabla \theta - {\bf A} ({\bf x}) \big\rvert ^2 + \Sigma ~P [{\cal D}] - \epsilon_{\rm bulk} ~ A [{\cal D}],
\end{equation}
where the first term describes the elastic cost of ``phase strain'' away from the locally preferred misfit of local orientation $\theta({\bf x})$.  The last two terms describe the respective cohesive cost and gain of the domain boundary and bulk (perimeter, $P [{\cal D}]$, and area, $A [{\cal D}]$) with respective line and bulk energies $\Sigma = J+K$ and $\epsilon_{\rm bulk} = 2\Sigma$.  As we summarize in Appendix \ref{mapping_appendix}, the elastic energy of this model can be directly related to the intrinsic frustration of in-plane orientational order (e.g. polar, nematic) on surfaces of non-zero Gaussian curvature $K_G$, connecting this generic model to a widely studied class of geometrically frustrated systems.  For frustrated 2D liquid crystalline order~\cite{nelson1987fluctuations, bowick}, the gauge field derives from the spin connection~\cite{kamien2002geometry} defined on a non-Euclidean surface for which $\nabla_\perp \times A = K_G$.  The microscopic connection between the uniformly frustrated XY model and orientational order on non-Euclidean surfaces is easily viewed in terms of the parallel transport of in-plane vectors in closed loops shown in Fig. \ref{fig: model}c-d, whose net precession grows with the integrated Gaussian curvature within the loops~\cite{needham_visual_2021}.  Additionally, recent work by Efrati and coworkers~\cite{NivEfrati, meiri2022XY} shows that distinct class of systems, planar bend-nematic phases, exhibits a variant of the 2D frustrated XY model, with degree of frustration controlled by the square of the preferred bend of the director (see Appendix \ref{mapping_appendix}).

Ground states of finite domain size $W$ adopt intra-domain gradient patterns in the phase strain that vary approximately linearly from boundary to boundary (Appendix \ref{continuum_appendix}), as shown for the square clusters in Fig.~\ref{fig: model}e.  These strain gradients lead to a mean elastic strain that grows as $\lvert \nabla \theta - {\bf A} ({\bf x}) \rvert \sim f W$, and super-extensive growth of the elastic energy density $\sim J f^2 W^2$.  At zero temperature, equilibrium between the competing effects of per subunit elastic and cohesive (boundary) energy of cluster formation selects a finite size
\begin{equation}\label{WStar}
W_* \approx \ell_{\rm d} \equiv \Big( \frac{ \Sigma/J }{f^2} \Big)^{1/3}
\end{equation}
Here $\ell_{\rm d}$ is a characteristic domain scale which, like the inter-vortex spacing, {\it decreases} with frustration and, contrary to the vortex-spacing, is dependent on the ratio of surface energy to elastic (spin) stiffness.  Note that this suggests defect-free domains approach the ``ferromagnetic'' state (i.e. $\nabla \theta-{\bf A}=0$) in the $f \to 0$ limit, with ground state phase-strain vanishing in the absence of frustration.  Although energy density diverges quadratically with domain size as $\sim Jf^2 W^2$ for finite frustration, the growth of optimal domain size $W_*$ as frustration grows is sufficiently slow as $f \to 0$ that the energy density due to frustration in these domains vanishes in this limit, albeit with a sublinear scaling $\sim J^{1/3}\Sigma^{2/3}f^{2/3}$.

\section{Self-Limitation at Intermediate Frustration}

We investigate the conditions for SLA at finite-temperature by way of canonical ensemble Monte Carlo simulation of our minimal model, over a wide parameter range (Appendix \ref{sim_appendix}). As we describe below, we find the key result that equilibrium SLA requires low ratios of cohesion to elastic stiffness, $\Sigma/J \ll 1$ (corresponding to the range $0<K<-J$).

\begin{figure}[ht]%
\centering
\includegraphics[width=0.5\textwidth]{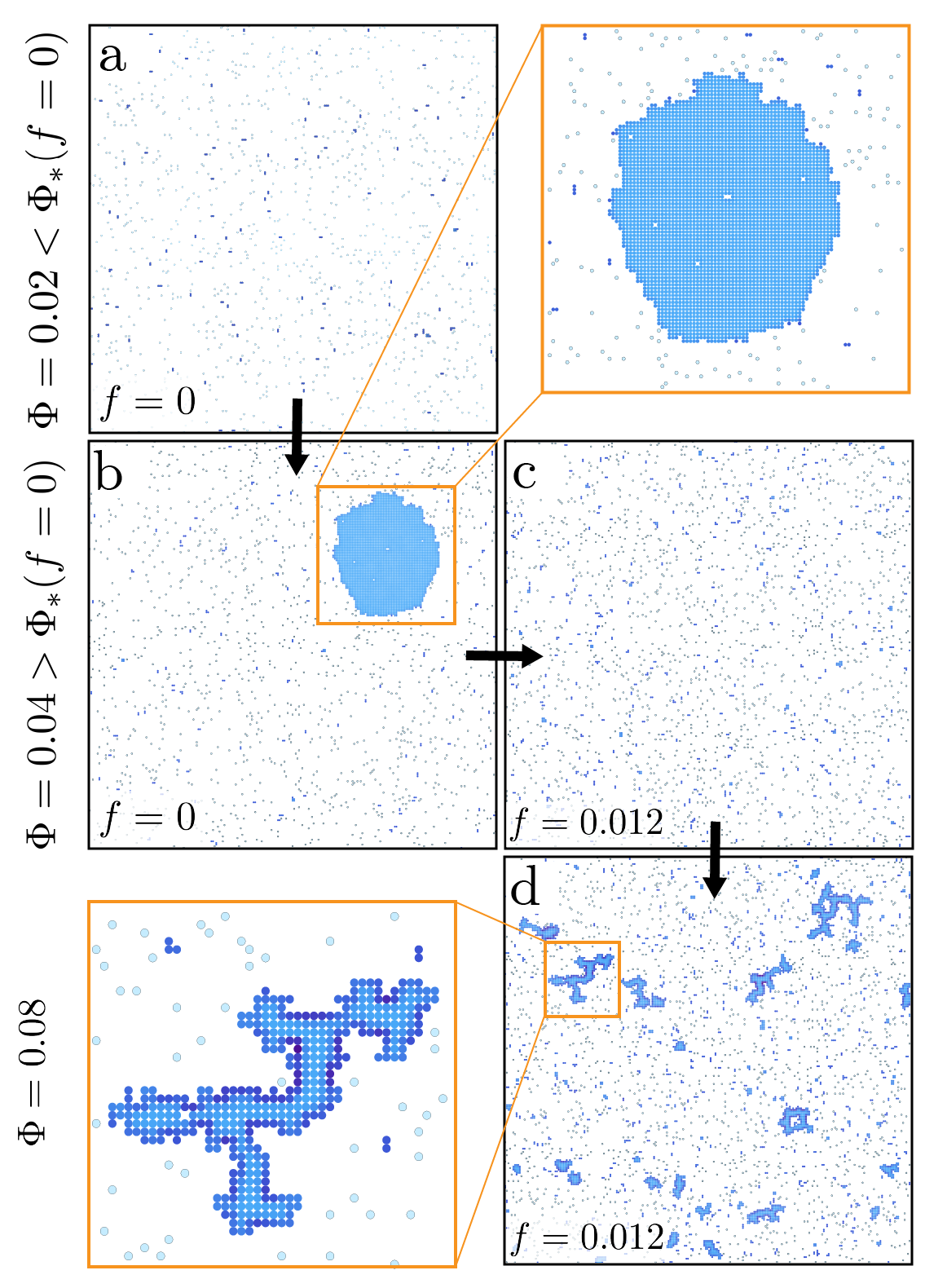}
\caption{\textbf{Zero-frustration condensation versus aggregation at strong frustration} Representative snapshots of equilibrated MC simulations of frustrated assemblies with $\Sigma/J=0.09$, $\beta J=40$ and \textbf{a} $f=0.0$, $\Phi=0.02$ (dispersed vapor) \textbf{b} $f=0.0$, $\Phi=0.04$ (vapor-droplet coexistence) \textbf{c} $f=0.012$, $\Phi=0.04$ (vapor) \textbf{d} $f=0.012$, $\Phi=0.08$ (finite-width aggregates). System size is $L=500$ and subunits are colored according to mean strain energy. }\label{fig: snapshots}
\end{figure}

We begin by illustrating the concentration dependence of the {\it unfrustrated} ($f=0$) model  for $\beta \Sigma = 0.09$ and $\beta J = 40.0$.  As shown in Fig.~\ref{fig: snapshots}a-b, for $\Phi<\Phi_*(f=0) \simeq 0.03$ the system is dispersed (\hyperlink{V1}{Supplementary Video 1}), and for $\Phi>\Phi_*(f=0)$ exhibits bulk phase separation between a vapor phase and a macroscopic droplet (\hyperlink{V2}{Supplementary Video 2}).  As these parameters are well below the expected temperature for spin ordering (i.e. the Kosterlitz-Thouless temperature $ \beta_{\rm KT} J \simeq 0.9$ \cite{ueda2021resolving}), the condensed droplets exhibit low values of phase strain, implying  ferromagnetic order.  Relative to the spin ordering transition, the system is much closer to the point of phase separation (i.e. the Ising critical point corresponding to $ \beta_{\rm Ising} \Sigma \simeq 1.8$ \cite{marko1995phase}), and therefore, droplets are round and exhibit visible capillary fluctuations.

Fig.~\ref{fig: snapshots}c shows a snapshot at the same concentration ($\Phi = 0.04$) but with a jump to strong frustration at $f = 0.012$, leading to {\it redispersal} into a vapor phase dominated by free subunits (\hyperlink{V3}{Supplementary Video 3}).  Aggregation at strong frustration requires pushing to even higher concentration values, as shown in Fig.~\ref{fig: snapshots}d at $\Phi = 0.08$ (\hyperlink{V4}{Supplementary Video 4}).  In this case, however, the state of assembly takes the distinct form of multiple aggregates coexisting with a population of free subunits.  As highlighted by the aggregate in Fig. \ref{fig: snapshots}d, at strong frustration these structures are distinct from $f=0$ droplets, due to their {\it anisotropic, wormlike} shapes and the {\it internal gradients of phase-strain}.  Although more complex in shape than the finite square domains in Fig \ref{fig: model}e, strongly frustrated wormlike aggregates exhibit self-limiting widths. This suggests a strong analogy between the minimal model of frustrated assembly and micellization~\cite{israelachvili_intermolecular_1992}, wherein the transition from low to high concentration takes the form of a pseudo-critical aggregation transition to a state of variable length, quasi-1D (i.e. finite width) clusters in equilibrium with a vapor of dispersed subunits.

\begin{figure*}[ht]%
\centering
\includegraphics[width=0.8\textwidth]{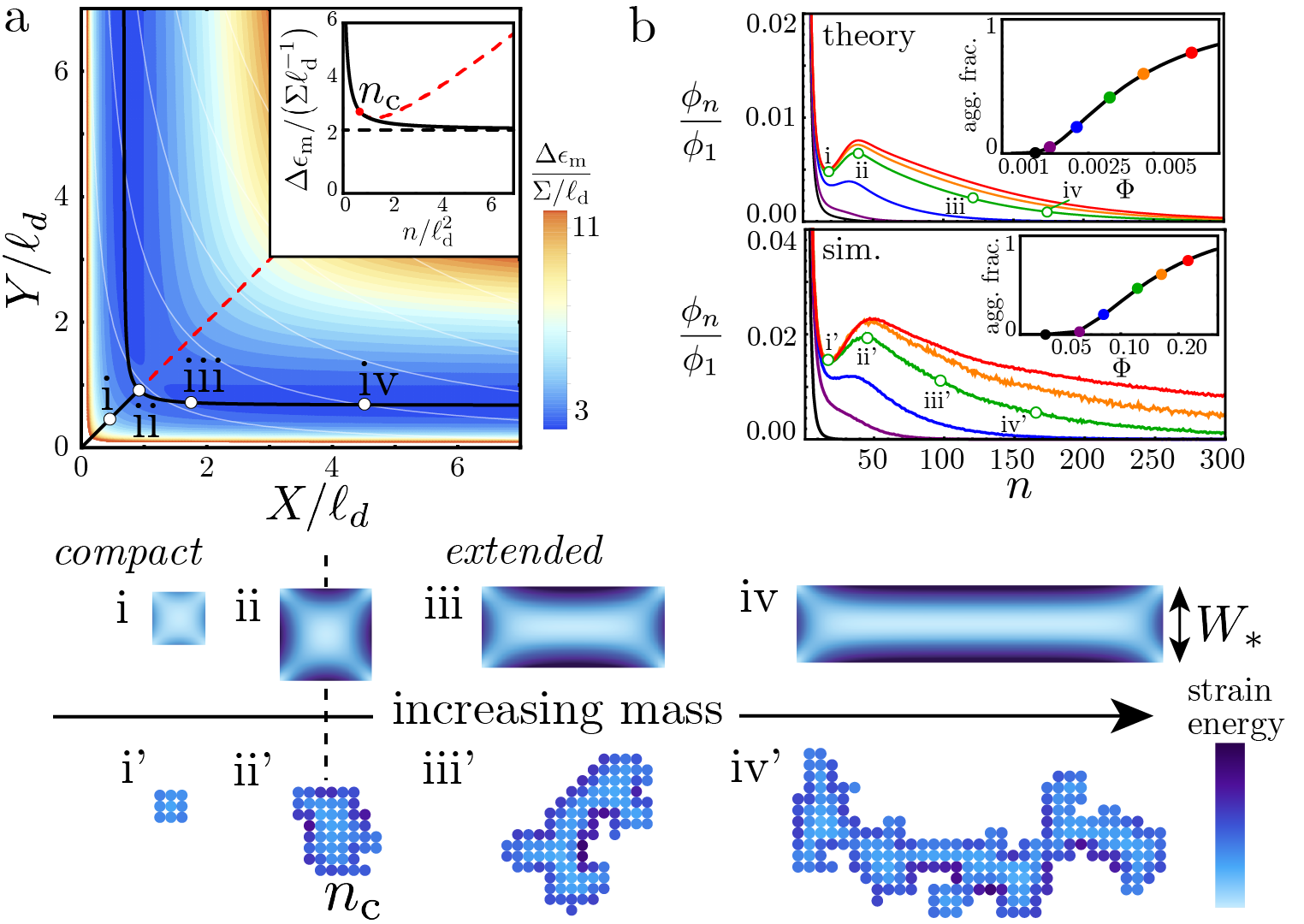}
\caption{\textbf{Self-limiting aggregation} \textbf{a} Continuum predictions of the energy density landscape as a function of lateral $x$ and $y$ dimensions of rectangular aggregates, with contours of  constant aggregation number ($n=xy$) shown in white. The minimal energy shapes as function of $n$ trace out the black path, and have a energy dependence shown in the inset, with dashed red curve showing the unstable portion of the square ($x=y$) branch.  The labeled points (i-iv) correspond to the aggregate structures shown below.  \textbf{b} Shows theoretical predictions (top) from the ideal aggregation of the continuum ``worm" model and MC simulation results (bottom) for mass fraction of $n$-mer aggregates for $f=0.016$, $\Sigma/J=0.09$ and $\beta J=40$.  The values of increasing concentration are indicated by color and correspond to the points highlighted in the inset plots of aggregated subunit fraction.  The same sequence of points (i-iv) are highlighted for theory (top), while similar $n$ values are chosen for the points (i'-iv') highlighted for the example states of lattice aggregates (bottom).} \label{fig: aggregation}
\end{figure*}


To understand the mechanism underlying the variable and anisotropic aggregate shapes, we consider the {\it energy per sub unit} $\epsilon$, of a simplified model of rectangular aggregates, modeled by ground states of the continuum energy in eq. (\ref{eq: continuum}) (Appendix \ref{continuum_appendix}).  Figure~\ref{fig: aggregation}a shows $\epsilon(X,Y)$, where $X$ and $Y$ are cross-sectional dimensions of the aggregate.  As aggregation thermodynamics is largely determined by the dependence of the interaction free energy of aggregates on subunit number $n=XY$, we focus on the optimal aggregate energy density $\epsilon_{\rm m}(n)$, which is minimized over aspect ratio $X/Y$ as a function of fixed $n$, Fig.~\ref{fig: aggregation}a inset.  The minimal-energy domain shapes, shown as black lines Fig.~\ref{fig: aggregation}a, are split into two regimes according to a critical aggregation number, $n_c\simeq0.837 \ell_{\rm d}^2$.  For small aggregates ($n \leq n_c$) optimal aggregates favor square shapes, Fig.~\ref{fig: aggregation}i-ii, while above this critical size ($n > n_c$), optimal aggregate shapes split off into two anisotropic rectangular branches which grow to arbitrary length, Fig.~\ref{fig: aggregation}iii-iv.  Emergent domain anisotropy is a feature of many models of frustrated assembly~\cite{schneider2005shapes, meng, hall2016morphology, witten, meiri2021cumulative}, in which optimal structures distribute strain gradients across their narrow dimensions to evade the super-extensive costs of frustration.  For large $n$, the optimal width of ``rigid worms" approaches a constant, self-limiting value $W_*(n \to \infty) = 3^{1/3} \pi^{-2/3} \ell_{\rm d}$, which unlike the thickness of micelles, may extend well beyond the subunit size. 

Similar to wormlike micelles~\cite{israelachvili_intermolecular_1992}, the energetic costs of ``end-caps'' favors arbitrarily long aggregates, yet the translational entropy favors equilibrium size distributions with a finite mean length, as opposed to bulk assembly. Fig.~\ref{fig: aggregation}b shows ideal aggregation theory predictions (Appendix \ref{continuum_appendix}) for the mass fraction $\phi_n$ of aggregates of size $n$ for an increasing sequence of total concentration $\Phi$ for the same strong frustration conditions simulated in Fig.~\ref{fig: snapshots}c-d. We predict a transition from a state dominated by free monomers, to one characterized by a secondary peak for $n \approx n_{\rm c}$ ($=42$ for this case) above a threshold concentration.  Because energetically optimal aggregates are infinite length (finite width) structures, we predict an exponential tail of $\phi_n$ with a width that grows with (square-root) concentration~\cite{HaganGrason}.
 \begin{figure}[h]%
\centering
\includegraphics[width=0.45\textwidth]{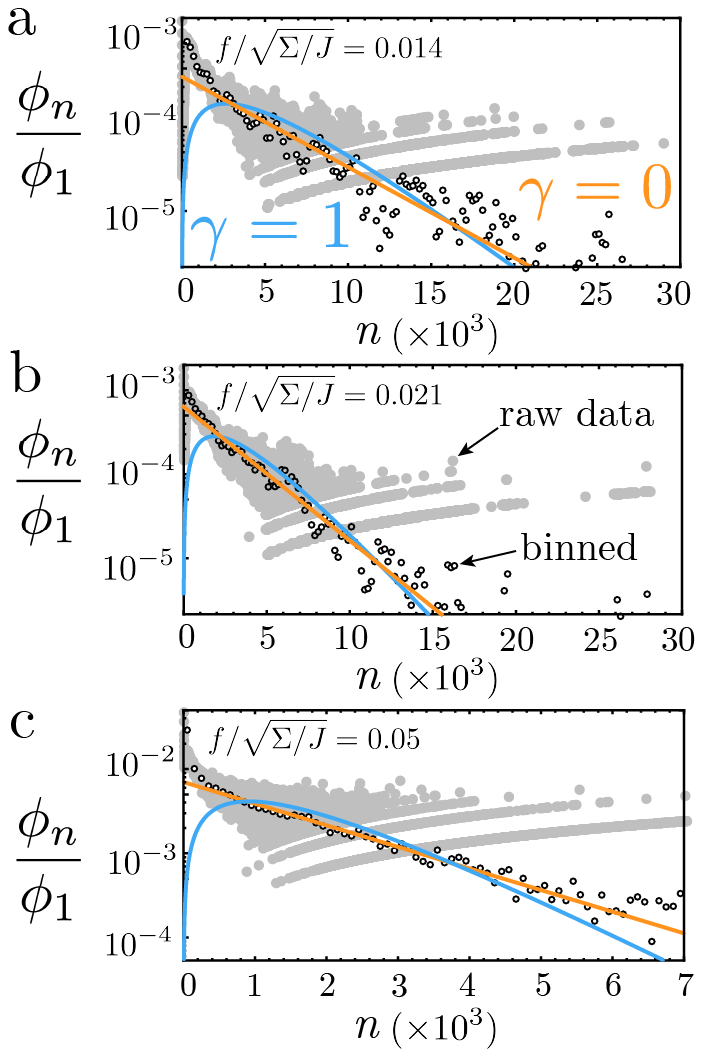}
\caption{\textbf{Statistics of large-$n$ aggregates} Large-$n$ tail of the aggregation concentration, measured for several different parameter sets: (\textbf{a}) $\Sigma/J=0.005$, $f=0.001$, $\beta J =650$ and $\Phi=0.3$; (\textbf{b}) $\Sigma/J=0.005$, $f=0.0015$, $\beta J =650$ and $\Phi=0.3$; and (\textbf{c}) $\Sigma/J=0.09$, $f=0.015$, $\beta J =40$ and $\Phi=0.3$). Numerical data is compared to the large-$n$ scaling $\phi_n \sim n^{\gamma} Z^{n}$. Best fits for both 2D branched polymer scaling~\cite{parisi1981critical} ($\gamma=0$, orange) and 1D linear chain aggregation~\cite{HaganGrason} ($\gamma=1$, blue) are compared. Fits were performed for $n>200$. Note that the largest aggregates in the raw numerical data (solid gray discs) occur only in small integer frequencies, leading to the visible spread between the $\phi_n\sim \ln n$ shaped bands in the distribution. In order to obtain a better view of the $\phi_n$ scaling in the large-$n$ limit, we binned the raw data into histograms with a bin width of $\Delta n= 200$ (black open circles). Fits shown in the figure were made using the binned data.}\label{fig: AggEntropy}
\end{figure}
 For comparison, Fig.~\ref{fig: aggregation}b shows the concentration dependence of the aggregation behavior of our lattice simulations.  Upon increasing $\Phi$ a transition from free monomers to aggregates occurs, peaked around $n\approx 50$, followed by tail that grows with concentration.  The crossover from monomer- to aggregate-dominated (aggregation fraction plots in the inset) is strongly shifted in concentration relative to the ideal aggregation rectangular clusters (continuum theory). We attribute this to undersampling of both lattice-scale configurational entropy as well as large-scale conformational fluctuations of the worm-like aggregate shapes of the continuum model.  Notably, the lattice aggregates show significant bend deformations, random branching and occasional looping, suggesting that their conformational statistics are likely in the universality class of branched polymers. Analysis of the large-$n$ tail of simulated aggregate distributions (Fig. \ref{fig: AggEntropy}) suggests a deviation from purely 1D aggregation statistics, characterized by $\phi_n (n \gg n_c) \sim n^\gamma Z^{n}$, where we find a value $\gamma \approx 0$ consistent with results for 2D branched polymers \cite{parisi1981critical}, and distinct from linear aggregation ($\gamma =1$)\cite{HaganGrason}. 
  \begin{figure}[h]%
\centering
\includegraphics[width=0.4\textwidth]{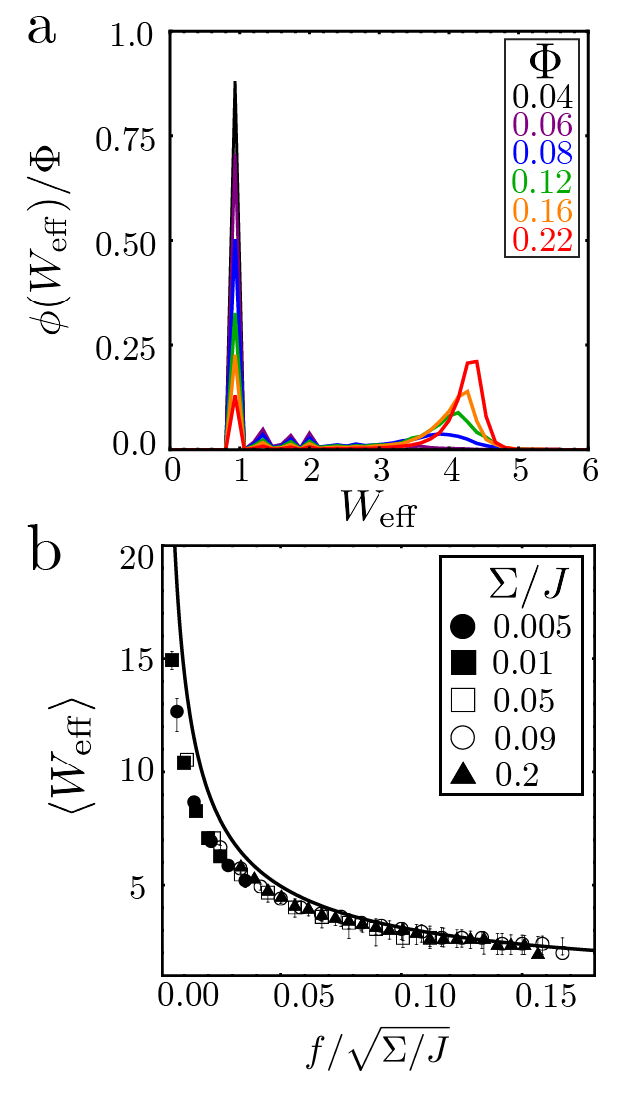}
\caption{\textbf{Self-limiting width selection} \textbf{a} Histograms of mass distribution as function of effective width of aggregated subunits for increasing values of total subunit concentration (same values of $\Phi$ as shown in Fig. \ref{fig: aggregation}b according to the color scheme). \textbf{b} Mean measured width $\langle W_{\rm eff} \rangle$ from aggregates in MC simulations for a range of $f$, $\Sigma/J$ and $\beta \Sigma$, compared to predicted dependence of the finite aggregate width (solid curve).  Error bars for $\langle W_{\rm eff} \rangle$ show the variance of width distribution. Full Parameters and corresponding symbols given in SI Table 3. }\label{fig: SLA Widths}
\end{figure}
 Despite the oversimplified model of fluctuations, the ideal continuum model captures the tightly regulated aggregate width. Figure ~\ref{fig: SLA Widths}a shows distribution of $W_{\rm eff}$, a measure of the mean subunit distance to the boundary (Appendix \ref{sim_appendix}), for the same parameters as Fig.~\ref{fig: aggregation}b , showing that aggregate widths fall into two distinct well-defined populations of free monomers and $W_{\rm eff} \simeq 4$ aggregates, independent of $\Phi$.  Figure~\ref{fig: SLA Widths}b compares the predicted self-limiting width $W_*$ for long-rectangular aggregates to $\langle W_{\rm eff} \rangle$ measured from simulations (supersaturated conditions) for a range of cohesion to stiffness ratios and strong frustration values, showing quantitative, fit-free agreement over nearly an order of magnitude of self-limiting dimension.  Beyond the gross dimension, analysis of the {\it shape} of simulated aggregates (Fig. \ref{fig: AggAnisotropy}) also shows the same transition from {\it compact} aggregates, for $n \lesssim n_c$, to {\it extended} (quasi-1D, finite-width) shapes for $n \gtrsim n_c$ as highlighted in Fig.~\ref{fig: aggregation}i?-iv?.  This mass-dependence of aggregate shape bears striking resemblance to the so-called ladder model of concentration driven transitions from spherical to cylindrical micelles~\cite{missel1980thermodynamic}.
 \begin{figure}[h]%
\centering
\includegraphics[width=0.425\textwidth]{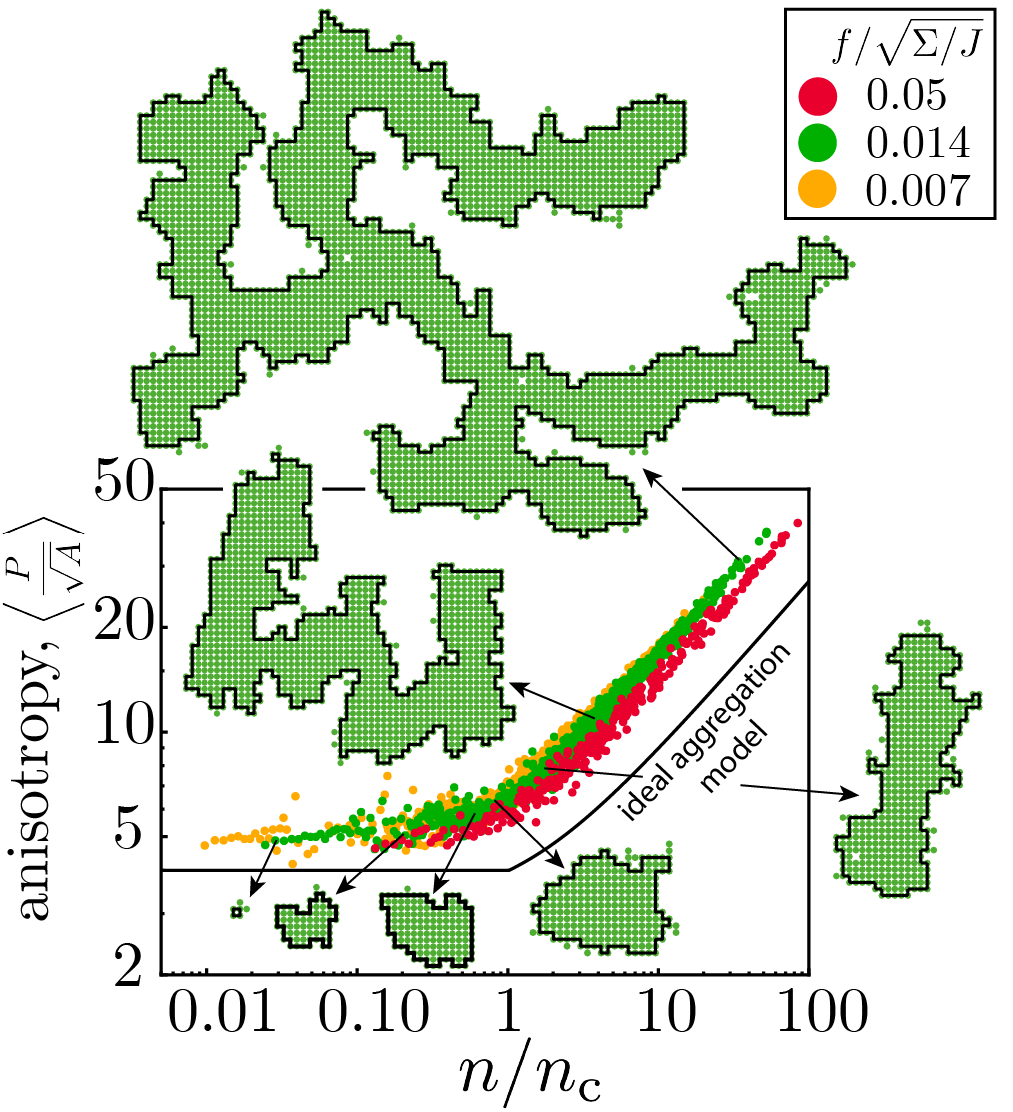}
\caption{\textbf{Compact to extended transition in frustrated assemblies} Ratio of outer cluster surface perimeter, $P$, to square root of enclosed area, $A$, (averaged over all observed aggregates) plotted as a function of $n$ over  $n_c$, the predicted critical size from continuum theory. Colors represent the following parameter sets: $\Sigma/J=0.09$, $f=0.015$, $\Phi=0.3$ (red); $\Sigma/J=0.005$, $f=0.0005$, $\Phi=0.3$ (orange); and $\Sigma/J=0.005$, $f=0.0010$, $\Phi=0.3$ (green). Example aggregates are taken from the data corresponding to the green parameter set. The solid black line corresponds to the ratio of $P$ to $\sqrt{A}$ calculated explicitly for aggregates living along the aspect ratio minimized ``saddle path'' (see Fig. \ref{fig: aggregation}a). The black line surrounding each example aggregate represents the domain boundary and the perimeter, $P$, for each aggregate is taken to be the length of this line. The aggregation area, $A$, is the number of subunits (green circles).}\label{fig: AggAnisotropy}
\end{figure}
The strong agreement between the continuum model and lattice simulations of frustrated aggregates, even under conditions where effects of conformational and configurational fluctuations are substantial, suggests that the purely energetic (elastic and cohesive) effects of the self-limiting ground states effectively govern key features of the free energy landscape of thermalized aggregates.  Further evidence of this is given by consideration of the critical aggregation concentration, $\Phi_*$, which characterizes the crossover from monomer- to aggregate-dominated states.  Elementary considerations of ideal aggregation theory for 1D aggregates imply that $\Phi_* \sim e^{\beta \epsilon_{\rm m} (n \to \infty)}$, which according to the continuum theory (Appendix \ref{continuum_appendix}) exhibits a generic dependence on the reduced frustration $f/\sqrt{\Sigma/J}$ the same dimensionless combination of frustration, cohesion and elastic stiffness that controls finite aggregate size.  Specifically, we find the simple relation, eq. (\ref{eq: CAC}), between the condensation point in the limit of zero frustration $\Phi_*(f=0)$ and the finite-$f$  aggregation concentration
\begin{equation}
\Phi_*(f) \approx \Phi_*(f=0)e^{C_0 \beta \Sigma  \big(\frac{f}{\sqrt{\Sigma/J}}\big)^{2/3}}  ,
\end{equation}    
where $C_0 = (3 \pi^2)^{2/3}/2$.  Accordingly Fig.~\ref{fig: SLAPhase}, demonstrates that this simplified analysis captures the key tendency for increased frustration to shift the aggregation point to larger $\Phi$ of a broad range of $\Sigma/J$ and temperature conditions.  
\begin{figure}[hb]%
\centering
\includegraphics[width=0.45\textwidth]{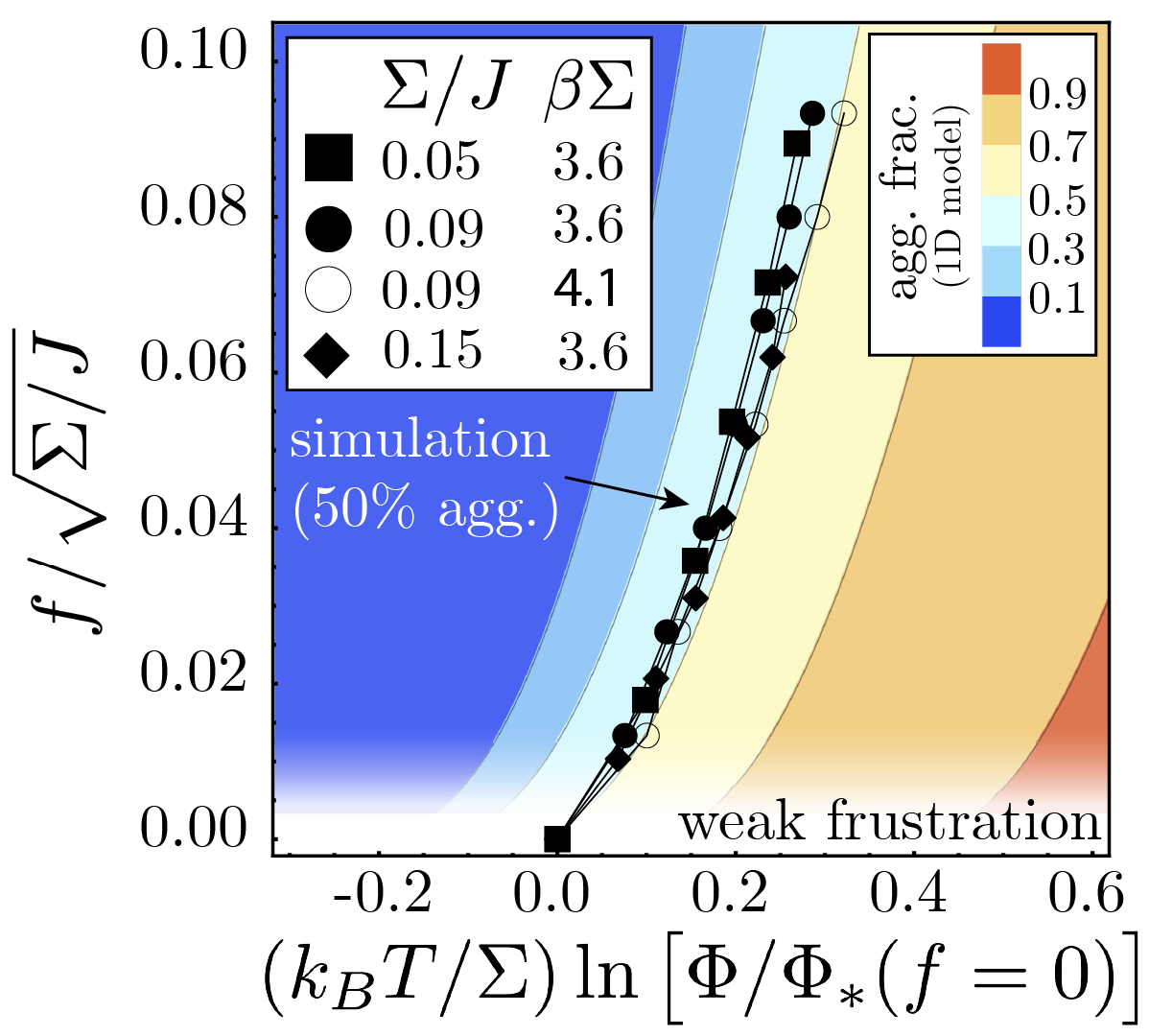}
\caption{\textbf{Pseudo-critical transition to self-limiting assembly} Comparison of aggregation conditions for simulations to continuum theory predictions, eq. (\ref{eq: CAC}), which give contours of increasing aggregation fraction as a function of scaled frustration and (log-)concentration, where $\Phi_*(f=0)$ corresponds to the onset of assembly (i.e. binodal) for the unfrustrated limit.  The points show where MC simulations reach 50\% aggregation.  The legend shows the values of $\Sigma/J$ and $\beta \Sigma$ corresponding to simulation results.}\label{fig: SLAPhase}
\end{figure}

\begin{figure*}[ht]%
\centering
\includegraphics[width=0.9\textwidth]{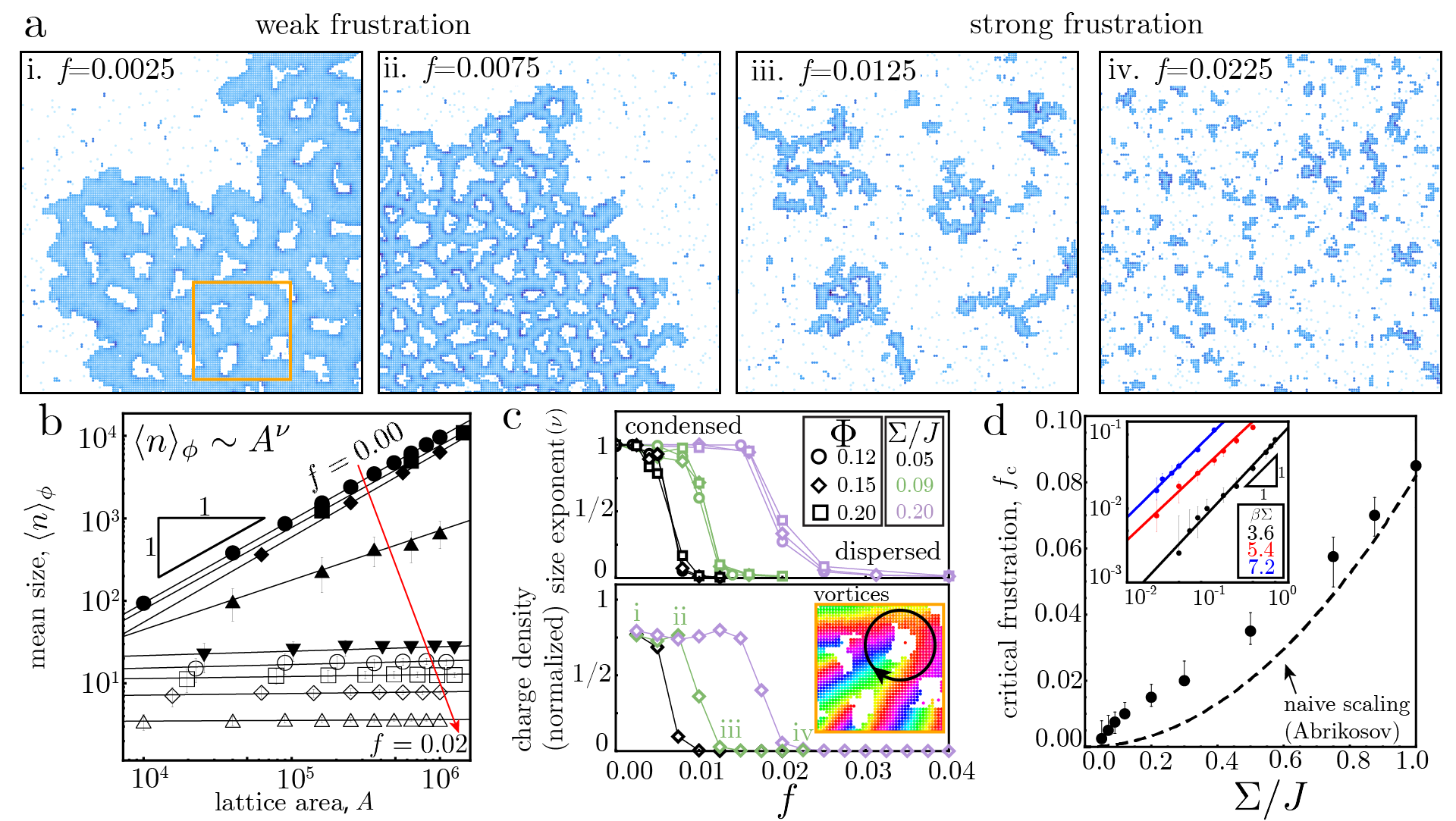}
\caption{\textbf{Condensation at weak frustration} \textbf{a} Simulation snapshots ($\Sigma/J=0.09$, $\beta J=40$, $\Phi=0.15$, $L=800$ and colored by strain energy) for a sequence of frustration values spanning the condensation transition. \textbf{b} Finite size scaling analysis of mean aggregate size for sequence of frustrations ranging from $f=0.0$ to $f=0.02$ (with $\Sigma/J=0.09$, $\beta J=40$ and $\Phi=0.15$). \textbf{c} size exponent (top panel) as a function of frustration extracted from finite size scaling analysis of simulations with $\Sigma/J=0.05, 0.09,0.20$ (black,green,purple) and $\Phi=0.12,0.15,0.20$ (open circles, diamonds, squares). The normalized charge density (bottom panel) -- measured as the net topological charge per subunit of an aggregate normalized by $f$ -- for the same values of $\Sigma/J$ (with $\Phi=0.15$), showing the coincidence of defect incorporation and bulk condensation and threshold frustration $_{\rm c}$ that is independent of $\Phi$. \textbf{d} The critical value of frustration as a function of $\Sigma/J$ from analysis of the mean size susceptibility (Appendix \ref{sim_appendix}) and compared to a naive (Abrikosov) scaling. Results for three different values of $\beta\Sigma$ (inset) reveal a temperature dependent shift in the critical frustration.}\label{fig: bulktodisperse}
\end{figure*}

\section{Condensation to Topologically-Defective Bulk}
A naive extension of the ideal aggregation theory described above would suggest a scenario in which worm-like aggregates exist for arbitrarily low frustration with finite-width $W_* \sim f^{-2/3}$ that diverges continuously as $f \to 0$.  Here we show, instead, that the transition from self-limiting aggregates at strong-frustration (Fig.~\ref{fig: snapshots}d) to the unfrustrated, bulk droplets (Fig.~\ref{fig: snapshots}b) is interrupted by a thermodynamic transition at a finite, critical value of frustration, $f_{\rm c}$, demarking the boundary between strong and weak frustration.  This transition is visible in the sequence in Fig. ~\ref{fig: bulktodisperse}a of simulations of variable frustration, spanning from weak to strong, where other parameters are held fixed.  The larger frustration values ($f \gtrsim 0.01$) correspond to self-limiting aggregate states, with finite widths that decrease with frustration.  At the lower frustration values ($f \lesssim 0.01$), however, assembled subunits adopt a completely different ``bulk sponge" morphology:  a macroscopic condensate, punctuated by an array of holes (see \hyperlink{V5}{Supplementary Video 5} for an example bulk sponge simulation).

Finite-size scaling analysis of the mean-cluster size, $\langle n \rangle_\phi$, with lattice area $A$, shown in Fig.~\ref{fig: bulktodisperse}b, confirms the weak-to-strong frustration transition as a thermodynamic transition between bulk condensation and dispersed phase of self-limiting aggregates.  The observed linear scaling for weak frustration, $\langle n \rangle_\phi \sim A$, is consistent with a mean-cluster size that is proportional to the macroscopic condensate, which occupies a fixed fraction of lattice area.  In contrast, for worm-like aggregates mean aggregate size is only a function of intensive quantities (e.g. $\Phi$, $T$, $f$) and is independent of system size, which is a well-known property of equilibrium quasi-1D assembly~\cite{HaganGrason}.  Hence, $\langle n \rangle_\phi\sim A^0$ for the dispersed states of self-limiting aggregates (as well as the disassembled vapor) at strong frustration.  

In Fig.~\ref{fig: bulktodisperse}c we analyze the dependence of frustration-driven condensation on concentration and cohesive to stiffness ratios. While the $f$-dependence of the finite-size scaling exponent $\langle n \rangle_\phi$ is independent of $\Phi$ (above saturation), the threshold frustration for the bulk-to-disperse transition clearly shifts to larger values of $f$ with increasing $\Sigma/J$.

Careful analysis of the phase-winding in the condensed sponges reveals that their mesoscopic holes are actually vortex defects (inset Fig.~\ref{fig: bulktodisperse}d).  The predominant topological charge of voided-vortices is $\pm 1 $ (with a sign opposite to $f$), although larger area holes, with higher net charge are observed (see Fig. S1).  Fig.~\ref{fig: bulktodisperse}d shows the mean charge density of aggregates normalized by $f$, the density of the ($\Phi = 1$) Abrikosov ground state, as a function of frustration.  Comparison to Fig.~\ref{fig: bulktodisperse}c illustrates that charged-to-neutral transition precisely coincides with the bulk-to-disperse transition.  Hence, underlying the condensation transition from strong-to-weak frustration is a transition in the primary mechanism for screening frustration, from free-boundaries of finite domains to extensive numbers of neutralizing vortices in macroscopic condensates.

These observations suggest a simple estimate for the boundary between strong and weak frustration, which follows from the assumptions that i) energetic costs of frustration dominate the thermodynamic competition between self-limiting and bulk states and ii) the bulk ``vortex sponge'' may be modeled by the $\Phi = 1$ Abrikosov lattice with a vortex spacing $\ell_{\rm v} \sim f^{-1/2}$~\cite{tarjus2005frustration}.  Comparing this length scale to the self-limiting scale $W_* \approx \ell_{\rm d} \sim f^{-2/3} (\Sigma/J)^{1/3}$, we expect finite-width domains to be much narrower than the inter-defect spacing ($\ell_{\rm d}\ll \ell_{\rm v}$) at large $f$, indicating the free-boundary formation is the favored mechanism of frustration screening.  As $f$ is lowered, finite domain sizes eventually exceed characteristic vortex spacing ($\ell_{\rm d}\gg \ell_{\rm v}$), suggesting that defect incorporation becomes favorable when these length scales crossover.  Using the Abrikosov scaling and the condition $\ell_{\rm v} \approx \ell_{\rm d}$ yields a naive estimate for the {\it critical frustration}
\begin{equation}
\label{eq: fc}
    f_{\rm c} \approx (\Sigma/J)^2 \  \ \ \ \ \ \ {\rm (Abrikosov \ scaling)}
\end{equation}
This scaling is qualitatively consistent with the increasing tendency of $f_{\rm c}$ with cohesion to elastic stiffness, as well as with the concentration independence, noted in Fig.~\ref{fig: bulktodisperse}d.  

\begin{figure*}[ht]%
\centering
\includegraphics[width=0.8\linewidth]{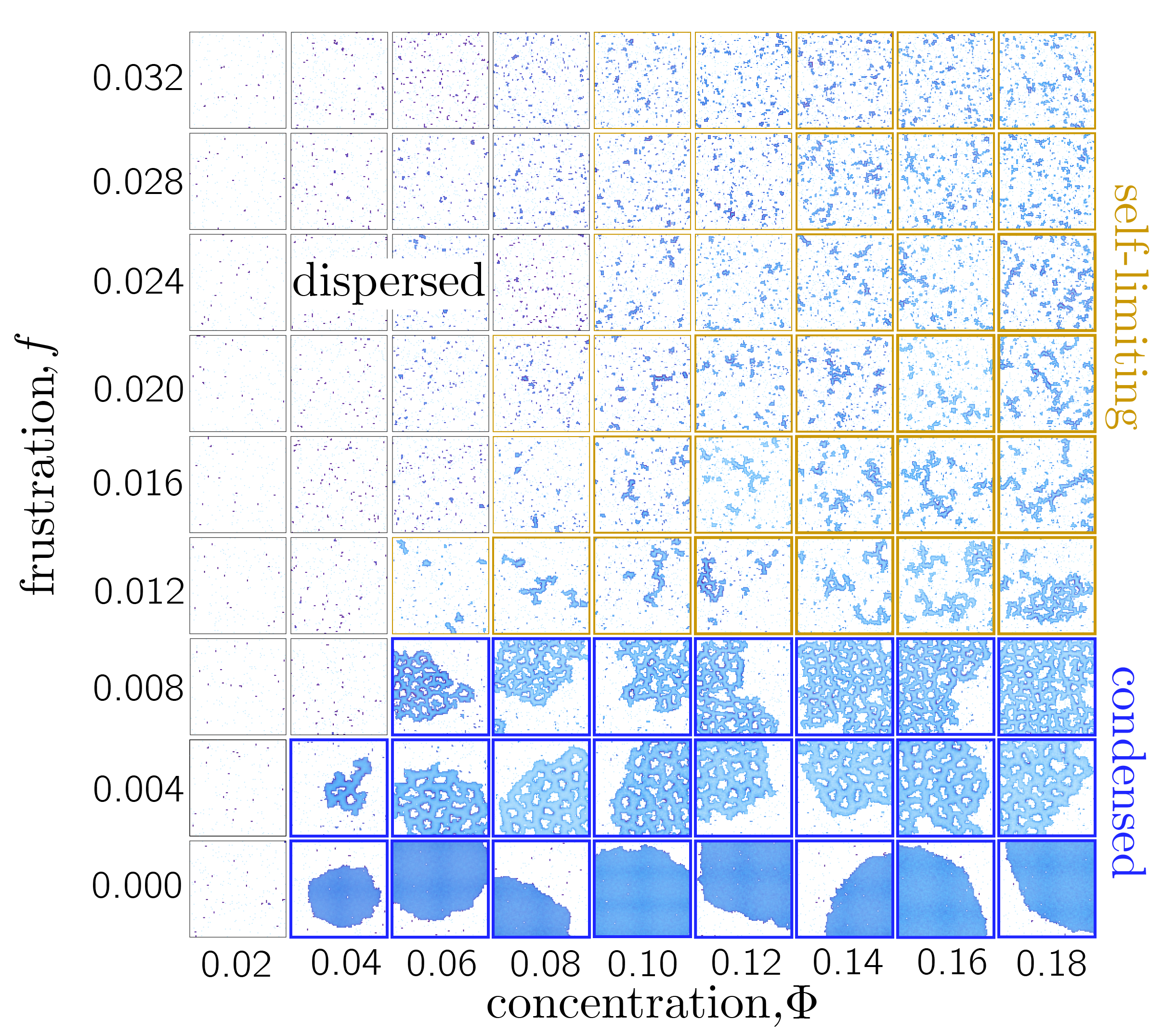}
\caption{\textbf{Phase diagram of frustrated assembly model} Mosaic of simulations in the $\Phi$-$f$ plane for $\Sigma/J=0.09$, $\beta J=40$ and lattice size $L=500$ (panels are $150 \times 150$ subsets). The panel boundaries are colored according to assembly state: (black) dispersed; (gold) self-limiting aggregates; and (blue) bulk condensates.  Line thickness of the self-limiting panels highlights three ranges of aggregation fraction:  (thin) $10-30\%$; (medium) $30-50\%$; and (bold) $>50\%$.  Subunits are colored according to mean (phase) strain energy as in Fig.~\ref{fig: model}.} \label{fig: phase}
\end{figure*}

Fig.~\ref{fig: bulktodisperse}e shows a critical analysis of the variation of $f_{\rm c}$ extracted from simulations over a broad range of $\Sigma/J$ values using a simpler fixed-area proxy for bulk-to-disperse transition (Appendix \ref{sim_appendix}).  While $f_{\rm c}$ indeed increases with $\Sigma/J$, the dependence is much closer to linear than the quadratic dependence predicted by the naive scaling argument in eq. (\ref{eq: fc}).  A clear limitation of this argument is neglect of the vortex core on the characteristic scaling which is evidently substantial in the = $W_* \gg 1$ regime.  Specifically, $\Sigma/J \ll 1$ implies relatively large core size (evident in Fig.~\ref{fig: bulktodisperse}a), perhaps generalizing the inter-vortex scaling to $\ell_{\rm v} \sim f^{-1/2} (\Sigma/J)^\nu$.  The linear scaling of $f_{\rm c}$ would be consistent with only a weak scaling with $\Sigma/J$ (i.e. $\nu \simeq 1/6$). While stronger dependence of defect spacing on $f$ is evident in bulk condensates, a putatively weaker dependence on $\Sigma/J$ is not possible to discern from simulations (see Appendix~\ref{sim_appendix}, Fig.~\ref{fig: Hole sep}).  

Lastly, simulations reveal a temperature-dependence in $f_{\rm c}$, which further conflicts with underlying assumptions and naive scaling for condensation in eq. (\ref{eq: fc}).  The inset of Fig.~\ref{fig: bulktodisperse}e plots the dependence of $f_{\rm c}$ on $\Sigma/J$ for three different temperatures, all of which are consistent with the scaling $f_{\rm c}\sim \Sigma/J$, but with a clear shift to higher frustration values as temperature is lowered.  Despite the clear evidence that cohesion and elastic cost of frustration control the self-limiting size (e.g. Fig.~\ref{fig: aggregation}), these results imply that the competition between self-limiting and bulk states is not purely energetic.  The evident decrease of $f_{\rm c}$ with $T$ further suggests that {\it entropy} of the finite aggregates plays a key role in their equilibrium stability at these temperatures.

\section{Phase behavior} Our minimal model exhibits {\it three distinct states} in a generalized phase diagram controlled by the combined temperature, concentration, frustration, cohesion and stiffness.  Fig.~\ref{fig: phase} shows the phase behavior in the $\Phi$-$f$ plane based on a mosaic of equilibrated simulation snapshots for fixed $\beta \Sigma = 0.09$, $\beta J = 40$.  While the vapor of dispersed monomers is stable at dilute concentrations, increasing $\Phi$ leads to two states of assembly dependent on the strength of $f$.  At weak frustration, the system phase separates when $\Phi$ crosses the binodal into coexistence between monomer vapor and defect-riddled bulk condensate (i.e. a vortex sponge phase). At strong frustration, increasing $\Phi$ drives aggregation into a dispersed state of finite-width, wormlike aggregates in equilibrium with a lower concentration of free monomers.  Over the dilute conditions studied here ($\Phi \lesssim 0.2$), the critical value of frustration for the self-limiting to bulk transition is independent of $\Phi$.  

From the mosaic of simulations in Fig. \ref{fig: phase}, one can see that\textemdash as frustration is lowered\textemdash the self-limiting aggregates become increasingly loopy, punctuated by holes that envelope vortices, as the condensation transition is approached. Despite this increase in loopiness, the SLA phase can still be clearly distinguished from the condensed bulk by the subunit mass distribution, which  shows that the SLA phase is a poly-disperse collection of aggregates with a finite size and a small number of holes per aggregate, while the condensed phase is a (quasi) bi-disperse phase with most of the subunit mass contained in a single bulk aggregate with a large number of holes. Analysis of the distribution of subunits by aggregate size and genus is provided in Fig. S1 for a range of frustrations spanning the condensation transition.

The tilt of the aggregation transition in the $\Phi$-$f$ plane for $f> f_{\rm c}$ leads to a generic sequence of phases at fixed saturated concentration above the $f=0$ binodal $\Phi*(f=0)$.  Notably, SLA is only stable at {\it intermediate} frustration, with very large values of $f$ driving dissolution of subunits while bulk condensation occurs for $f<f_{\rm c}$.

\section{Discussion}

\subsection{Critical behavior and connected models}
\label{sec: connected}
The minimal model of frustrated assembly can be situated between two well-studied physical models, each of which represent a limiting regime of eq. (\ref{eq: Hamiltonian}) and exhibit unique phase transition behaviors. The fully occupied limit $\Phi \to 1$ corresponds to the uniformly frustrated 2D XY model, which has been studied as a model of field-frustrated superconducting arrays~\cite{teitelprl1983,teitelprb1983,franz1995vortex} as well as a statistical framework for frustration models of glasses~\cite{tarjus2005frustration, Esterlis}.  Bulk states incorporate a finite density of vortices for any $0<|f|\ll 1/2$, leading to a so-called avoided critical point in the $f\to 0$ limit of the fully occupied model~\cite{KIVELSON199527, CHAYES1996129}.  In this scenario, ``melting'' of spin order by the Kosterlitz-Thouless (KT) mechanism is preempted by unbound vortex order at much lower temperature.  Alternatively, the limit of $f=0$ but {\it variable} concentration corresponds to the ``vectorized" Blume-Emory-Griffiths (BEG) model~\cite{berker} -- a 2D XY variant of the canonical BEG model~\cite{beg} -- that describes critical behavior in phase-separating superfluid films.  This model describes the interplay between spin ordering and condensation (i.e. translational order) and is marked by a line of spin-ordering transitions at high $\Phi$ that descend from the standard ($\Phi =1$) KT critical temperature towards the Ising-like binodal region describing phase separation.  As the ratio of spin stiffness to Ising interactions increases, the intersection of spin-ordering and binodal curves approaches the Ising critical point, resulting in (nearly) tricritical behavior.

Equilibrium SLA behavior exists at the intermediate regime to these limiting scenarios, i.e. finite $f$ and $0<\Phi<1$, yet the precise nature of its phase transitions remains to be understood.  One clue comes from the pronounced restructuring of concentration fluctuations by frustration.  Relative to spinodal or droplet-like fluctuations, at finite (strong) frustration, concentration fluctuations take the form of highly-anisotropic, string-like aggregates, effectively living polymers, and it is natural to expect these to strongly alter the transition to the high-density (bulk) phase.  For example, at higher $\Phi$ we anticipate that aggregate interactions become significant, and likely exhibit a percolation transition to a bulk state.  The randomly branched polymer structure of these aggregates suggests further that even at lower $\Phi$ the disperse-to-bulk transition may be described as a type of sol-gel transition~\cite{stauffer}.  This scenario is consistent with the evident increase of $f_{\rm c}$ with temperature, which itself suggests that {\it conformational entropy} of finite-width aggregates stabilizes their disperse state relative to the gel-like, condensed state.  It remains to be understood how the interplay of $\Sigma, J$ and $f$ determine the shape energetics of conformational fluctuations (e.g. aggregate bending, branching and looping).

\subsection{Experimental implications}

The results of the model presented here hold more general lessons for a broad class of frustrated assemblies where the frustration leads to quadratic power-law accumulation of elastic energy with size, including but not limited to the examples of frustrated orientational order described in Appendix \ref{mapping_appendix}.  In addition to the predicted thermodynamic stability of the self-limiting state at an intermediate frustration regime, the model also constrains the parameter regime where finite sizes can be non-trivially larger than individual subunits, i.e. where $W_* \gtrsim 1$. Furthermore, as the maximal self-limiting dimensions occur at the boundary with the bulk state, we estimate the upper size limit as $W_*(f_{\rm c}) \sim (\Sigma/J)^{-1/3}$, using the linear scaling of $f_{\rm c}$ from Fig.~\ref{fig: bulktodisperse}e. This suggests that non-trivial finite sizes require high stiffness to cohesion ratios $J \gg \Sigma$.  Since subunit binding requires temperature at or below $T_{\rm Ising} \approx \Sigma$, this further implies that SLA only occurs well below the nominal KT temperature\textemdash deep in the phase-ordered state\textemdash suggesting that elastic fluctuations play little, if any, role in SLA behavior.

Our minimal model of frustrated assembly applies most directly to 2D liquid-crystalline assemblies embedded on frozen non-Euclidean templates, where Gaussian curvature and hence frustration is fixed. The best studied examples of this include nematically ordered assemblies of rod-like particles on spheres~\cite{smallenburg2016, testa2018self} or spherical shells of nematic liquid-crystals prepared via double-emulsion droplets with planar anchoring~\cite{Nieves2007, lopez2011frustrated, lopez2009}.  The prior focus of studies on this latter class of systems has focused exclusively on complex defect states exhibited by ``bulk'' shells, where nematics cover the entire surface~\cite{lopez2011drops}.  In this system, the bulk to self-limiting transition would occur in mixed shells composed of controlled fractions of the nematic and isotropic phase, for example as would occur in shells of lyotropic liquid crystals.  In this case, line tension between these phases determines $\Sigma$, the elastic stiffness is controlled by the Frank elastic constant $K$, while the frustration is Gaussian curvature -- the inverse square of the shell radius.  Predictions of the lattice model of frustrated assembly then suggest that finite width self-limiting domains occur on small shells, while the nematic domains condense into a single, defective bulk on large shells, with a transition occuring at a critical radius set by temperature and the ratio of line tension to Frank constant.  It is conceivable that fixed frustration studies in thin nematic/isotropic layers could be extended to negative Gaussian curvature, for example by infusion into mesoporous minimal-surface-like structures that introduce an additional feature of spatially varying, yet quenched, frustration to the problem. These structures could be 3D-printed using techniques similar to~\cite{Serra2011}.

In Appendix~\ref{mapping_appendix} we detail the mapping of the elastic energy of two additional classes of systems onto the frustrated XY model of the form given in eq. (\ref{eq: continuum}). Tilted, hexatic, or otherwise liquid crystal phases of surfactant membranes are well-studied for the interplay between the in-plane order and the preferred out-of-plane shapes~\cite{oda1999tuning, Selinger1996, selinger2001theory}. Here, generic principles imply that liquid crystalline order of chiral surfactants induces a preference for membrane shapes with negative Gaussian curvature~\cite{Helfrich1988} of the form that frustrates their lateral assembly~\cite{GhafBruin, armon2014shape, Sharon}.  Additionally, recent work by Efrati and coworkers~\cite{NivEfrati, meiri2022XY} has shown that the planar bend-nematic liquid crystalline phase exhibits a variant of frustrated 2D XY order, and as such, assemblies of curved molecules or particles will lead to the super-extensive growth of elastic energy. Here, the effective frustration is determined by the {\it preferred bend} of the director field.  Indeed, recent advances in the fabrication and study of banana shaped colloids~\cite{kotni2022, fernandez2020shaping, fernandez2021synthesis} has led to the exploration of planar assembly of bulk states exhibiting complex arrays of vortices and other defects that resolve the cost of frustration~\cite{fernandez2021hierarchical}.  Extending these studies to lower area fractions of cohesively assembling particles, via e.g. depletion induced attraction, opens up the possibility to investigate self-limiting states and the condensation transition. It should be noted (see Appendix ~\ref{mapping_appendix}) that these two examples are critically distinct from the case of fixed frustration, say for spherical nematics, since the gauge field, and hence the source of frustration, is free to relax in equilibrium, alongside the assembly process itself. This introduces new questions about the potential transition to alternative bulk states, a point which we return to below.

Beyond these models for which the mapping onto the specific frustrated 2D XY model is known, we anticipate that predictions of our generic model will hold for a broader class of ?ill fitting? particle assemblies. Namely, there have been several recent and ongoing efforts focused on learning how to translate the mechanism of self-limitation in frustrated systems into a design principle for size-programmable assembly, for example\cite{witten,serafin,spivack2022stress,tyukodi2022thermodynamic,Berengut2020,Hall2023,tanjeem2022}. Notwithstanding differences in the microscopic mechanisms of frustration and deformation, many of these models are shown to exhibit the same (quadratic) power-law growth of elastic energy with finite domain size. For example, it can be shown that, in the limit of the narrow sizes where bending elasticity dominates, the elastic energy dependence of so-called warped jigsaw particles exhibits identical dependencies on the shape and size of domains\cite{spivack2022stress}. We anticipate, therefore, that any such 2D frustrated assembly exhibiting the same power-law scaling, will be described by the assembly behavior of our generic model, particularly in the strong-frustration regime where domain sizes are sufficiently small that this generic power-law dependence holds.

One generic ingredient of frustrated assembly behavior is the elasticity of frustration itself, which is absent from the present model wherein $f$ is fixed (akin to a frozen topography).  In many physical cases, there is a finite elastic cost to {\it flatten}, or defrustrate, the system~\cite{HaganGrason}.  For frustrated liquid crystalline membranes this takes the form of a finite elastic cost to ``flatten'' away from the preferred non-Euclidean shape~\cite{armon2014shape, Sharon, Hall2023}, while for bend-nematic assemblies this is controlled by the finite elastic cost of deviation from the preferred bend~\cite{NivEfrati, meiri2022XY}.  Hence, in addition to the defect-riddled bulk, the shape elasticity of the frustration source gives rise to alternative mechanisms of ``escaping" frustration to a defrustrated bulk state.  We anticipate that incorporation of gauge elasticity into the minimal model will give rise to yet even richer thermodynamics, which are controlled by an additional ``shape-flattening'' length that plays a role analogous to the penetration length in superconductivity and is likewise populated by two distinct bulk states, the analogs of the type-I and type-II superconducting phases.

\section{Acknowledgements}
The authors are grateful to J. Machta for numerous discussions on MC simulation approaches to the lattice model, to  R. Kamien, M. Wang and M. Hagan for valuable input on this manuscript, and to T. Lopez-Leon, A. Fernandez-Nieves and F. Serra for insight into experiments on curvature-frustrated liquid crystals.  This work was supported by US National Science Foundation through award NSF DMR-2028885.  Simulations where performed using the UMass Cluster at the Massachusetts Green High Performance Computing Center.

\section{Supplementary Information}

\noindent {\bf Supplementary Information} - Supplementary Figure 1, Supplementary Tables 1-8 and supporting notes.

\noindent \hypertarget{V1}{{\bf Supplementary Video 1}} (\url{http://www.pse.umass.edu/sites/default/files/grason/images/unfrustrated\_dispersed.mp4})) - Animation of MC simulations of unfrustrated disperse phase, shown in Fig.~\ref{fig: snapshots}a:  $\beta J = 40$, $\Sigma/J = 0.09$, $f=0$ and $\Phi = 0.02$.  For purposes of visualization, global cluster moves are turned off at the later stage of the video.


\noindent \hypertarget{V2}{{\bf Supplementary Video 2}} ( \url{http://www.pse.umass.edu/sites/default/files/grason/images/droplet_phase.mp4})- Animation of MC simulations of unfrustrated condensed droplets, shown in Fig.~\ref{fig: snapshots}b:  $\beta J = 40$, $\Sigma/J = 0.09$, $f=0$ and $\Phi = 0.04$.  For purposes of visualization, global cluster moves are turned off at the later stage of the video.


\noindent \hypertarget{V3}{{\bf Supplementary Video 3}}  (\url{http://www.pse.umass.edu/sites/default/files/grason/images/frustrated_dispersed_phase.mp4}) - Animation of MC simulations of strong frustration, disperse phase, shown in Fig.~\ref{fig: snapshots}c:  $\beta J = 40$, $\Sigma/J = 0.09$, $f=0.012$ and $\Phi = 0.04$.  For purposes of visualization, global cluster moves are turned off at the later stage of the video. 


\noindent \hypertarget{V4}{{\bf Supplementary Video 4}}   (\url{http://www.pse.umass.edu/sites/default/files/grason/images/self-limiting_phase.mp4}) - Animation of MC simulations of self-limiting aggregation at strong frustration, shown in Fig.~\ref{fig: snapshots}d:  $\beta J = 40$, $\Sigma/J = 0.09$, $f=0.012$ and $\Phi = 0.08$.  For purposes of visualization, global cluster moves are turned off at the later stage of the video.


\noindent \hypertarget{V5}{{\bf Supplementary Video 5}}  (condensed\_phase.mp4, \url{http://www.pse.umass.edu/sites/default/files/grason/images/condensed_phase.mp4}) - Animation of MC simulations of condensed ``vortex sponge'' formation at weak frustration:  $\beta J = 40$, $\Sigma/J = 0.09$, $f=0.004$ and $\Phi = 0.06$.  For purposes of visualization, global cluster moves are turned off at the later stage of the video.

\appendix

\section{Simulations \label{sim_appendix}}
\renewcommand{\thesubsection}{A.\arabic{subsection}}
\noindent {\bf Lattice Geometry and Gauge Field} We consider partially occupied $L\times L=A$ square lattices with periodic boundary conditions and fixed number of subunits $N\equiv\sum_i \eta_i = \Phi A$, where $\eta_i=0$ or 1 is the occupation at site $i$. 

We consider a divergence-free, uniformly frustrated gauge field,
\begin{equation}
{\bf A}({\bf x}) = \pi f ~ \big[ y \hat{x} - x \hat{y} \big]
\end{equation}
where lengths in ${\bf x} = (x,y)$ coordinates are measured in units of the lattice spacing.  This gives the gauge variables on nearest neighbor bonds,
\begin{equation}
A_{ij}=\pi f  ({\bf x}_i \times {\bf x}_j) =\left\{ \begin{array}{ll} \pi f y, & \ ij \parallel \hat{x} \\ - \pi f x, & \ ij \parallel \hat{y} \end{array} \right. 
\end{equation}
where ${\bf a} \times {\bf b}=a_x b_y-a_y b_x$. Periodic boundary conditions introduce a gauge-field requirement that Wilson loops wrapping around the periodic directions of the simulation box (i.e. the in $\hat{x}$ and $\hat{y}$)\cite{alba2008uniformly} take on the specific value: 
\begin{equation}
\mathcal{W}_{\hat{x}/\hat{y} }=e^{i\sum_{\mathcal{C}_{\hat{x}/\hat{y} }}A_{ij}}=1
\end{equation}
where $\mathcal{C}_{\hat{x}/\hat{y} }$ is a closed loop spanning the lattice in the  $\hat{x}$ or $\hat{y}$ direction. We satisfy these conditions on the $L\times L$ lattice by restricting to values of frustration of the form $f=2m/L$ where $m$ is an integer.

\noindent {\bf Markov chain Monte Carlo} We performed simulations at a fixed concentration, $\Phi,$ and temperature, sampling the Boltzmann distribution via an implementation of the Metropolis-Hastings algorithm.  The procedure is organized into a set of {\it sweeps}, which take a given state of an $L\times L$ lattice variables to a new state via a set of trial moves for positions and phases of occupied subunits, which we also denote as ``particles".  Trial moves are of two types:  {\it multi- particle cluster} (MPC) and {\it single-particle} (SP).  Lattice sites are indexed by integer $i$.  For generating MC moves at fixed concentration it is convenient to index occupied subunit positions (i.e. sites for which $\eta_i=1$) via the label $p=1\ldots N$, and identify the lattice position of $p^{\rm th} $ particle as $i(p)$, and track only spin variables of the particles (i.e. occupied sites), which we denote as $\theta_p$.  In practice, spin variables for {\it unoccupied} positions (i.e. where $\eta_i=0$) are not considered.

In each sweep $s$ the following sequence of trial moves are included:
\indent \begin{enumerate}
    \item {\it Cluster inversion} (one trial)
    \item {\it Cluster Wolff rotation } (one trial)
    \item {\it Single-particle moves} ($N$ iterations)
    \begin{enumerate}
    \item[3.a] {\it Local translation} (one trial)
    \item[3.b] {\it Global translation} (one trial)
    \item[3.c] {\it Phase rotation} (one trial)
    \item[3.d] {\it Rotation + translation} (one trial)
    \end{enumerate}
\end{enumerate}
where we summarize each move type below.  A typical simulation is run for $N_s =10^5-6 \cdot 10^6$ sweeps with configurations sampled every $10^2$ to $10^3$ sweeps.  This sampling interval is chosen to exceed the measured auto-correlation times of cluster relaxation in the self-limiting aggregate state (see below). Parameter values (including lattice size) are given SI Tables S1-11 for all analyzed simulations.    

\noindent {\bf Single-particle moves} We first describe the SP moves, all of which are accepted with a probability 
$P=\min(1,e^{-\beta\Delta E})$ where $\Delta E$ is the energy difference between the trial and initial state.

\begin{description} 
\item [Local translation]
\begin{enumerate}
\item  [i]Select random particle $p$
\item  [ii]Choose random {\it nearest neighbor} site $j$ to $I(p)$ 
\item  [iii]If neighbor is vacant (i.e. if $\eta_j = 0$), then displace $p$ to neighbor position:  $i(p) \to j$
\end{enumerate}

\item [Global translation]
\begin{enumerate}
\item [i] Select random particle $p$
\item [ii] Choose random {\it unoccupied} site $j$  
\item [iii] Displace $p$ to site $j$:   $i(p) \to j$
\end{enumerate}

\item [Phase rotation]
\begin{enumerate}
\item [i]Select random particle $p$
\item [ii]Choose random rotation $\delta \theta\in [ -0.1, 0.1]$
\item [iii]Rotate phase: $\theta_p\to \theta_p + \delta \theta$
\end{enumerate}

\item [Rotation/translation]
\begin{enumerate}
\item [i]Select random particle $p$
\item [ii] Choose random {\it unoccupied} site $j$  
\item [iii]Choose random rotation $\delta \theta\in [ 0, 2 \pi]$
\item [iv]Rotate (particle) phase : $\theta_p\to \theta_(p)+ \delta \theta$
\item [v]Displace (rotated) $p$ to site $j$:   $i(p) \to j$.
\end{enumerate}

\end{description}

\noindent {\bf Cluster moves } While self-limiting aggregates rapidly equilibrate with SP moves alone, the formation of defective (vortex sponge) bulk structures is subject to kinetic trapping into metastable states and requires additional non-local cluster moves to facilitate equilibration. To this end, we designed a Wolff-like \cite{Wolff1989cluster} {\it cluster inversion} move that is a modification of one described by Liu and Luijten in \cite{Luijten2004cluster} and Dress and Krauth in \cite{dress1995cluster} to equilibrate a simple liquid composed of hard disks. The move is rejection free and works by point reflecting a cluster of particles about a randomly chosen reflection point, ${\bf r}$. The new coordinates of a reflected particle (i.e. point inverted) are given by
\begin{align}
\label{eq: reflection}
     (x, y) \to ({\rm mod}[2 r_x - x, L], {\rm mod}[2r_y -y, L]),
\end{align}
where $(x,y)$ are the initial coordinates and the modulo operator is taken with respect to lattice size.

The presence of a gauge field makes it so that simply moving the position of a cluster changes its energy and internal strain.  For two sites $i$ and $j$ undergoing the point reflection in eq. (\ref{eq: reflection}), it is straight forward to show that the gauge field on the $ij$ bond transforms as $A_{ij} \to A_{ij} - 2 \pi f ({\bf x}_i-{\bf x}_j) \times {\bf r}$.  Therefore, to generate moves that are isoenergetic with respect to phase strain {\it within the inverted cluster} it is necessary to ``parallel transport'' the spins according to
\begin{align}
\label{eq: paralleltrans}
    \theta_p \to \theta_p - 2 \pi f ({\bf x}_{i(p)} \times {\bf r}) + \delta \theta,
\end{align}
where $\delta \theta$ is a randomly chosen (constant) phase rotation performed on the entire cluster. 

The outline of the algorithm, as shown below, is very similar to that in Ref.\ \cite{Luijten2004cluster} with a few small modifications.
\begin{description}
    \item[Cluster inversion]
    \begin{itemize}
        \item[i] Choose random reflection point ${\bf r}$
        \item[ii] Choose random global phase rotation $\delta \theta\in [ 0, 2 \pi]$
        \item[iii] Choose random particle $p$ and point reflect position and phase according to eqs. (\ref{eq: reflection}) and (\ref{eq: paralleltrans}).  If the new location is occupied, swap particles, parallel transporting both.
         \item[iv] Iterate  following sequence after each inversion until no possible particles can be added to the inverted cluster:   
\begin{enumerate}
    \item[a] Add to list of neighboring particles to updated sites and the changes in pair bond energy, $\Delta E_b$, after a proposed point reflection or double-reflection (i.e. when reflecting to occupied sites) about ${\bf r}$ + parallel transport
    \item[b] Choose a neighbor from the list to reflect + parallel transport, and accept that move with probability ${\rm min}(1, 1-\exp[-\beta \Delta E_b])$. Remove considered particle from list of cluster neighbors.
    \item[c] Particles can only be reflected once, but may be added to the neighbor list multiple times (once per cluster bond).
\end{enumerate}
  \item[(iv)] Cluster move is completed when neighbor list is empty.
    \end{itemize}
\end{description}

In addition to the cluster inversion, we perform a standard Wolff rotation\cite{Wolff1989cluster} on a cluster of occupied sites, which can be summarized as follows. 
\begin{description}
    \item[Cluster Wolff rotation]
    \begin{itemize}
        \item[i] Choose random global phase rotation $\delta \theta\in [ 0, 2 \pi]$
        \item[ii] Choose random particle $p$ and rotate its spin by $\theta_p \to \theta_p + \delta \theta$
         \item[iii] Iterate following sequence after each rotation until no possible particles can be added to the rotated cluster:   
\begin{enumerate}
    \item[a] Add to list of neighbor particles and compute the change in pair bond energy, $\Delta E_b$, after a proposed phase rotation by same $\delta \theta$
    \item[b] Choose a neighbor from the list to rotate, and accept its rotation with probability ${\rm min}(1, 1-\exp[-\beta \Delta E_b])$. Remove considered particle from list of cluster neighbors.
    \item[c] Particles can be rotated only once, but may be added to the neighbor list multiple times (once per cluster bond).
\end{enumerate}
  \item[(iv)] Cluster move is completed when neighbor list is empty.
    \end{itemize}
\end{description}

\begin{figure*}
\centering
\includegraphics[width=0.8\linewidth]{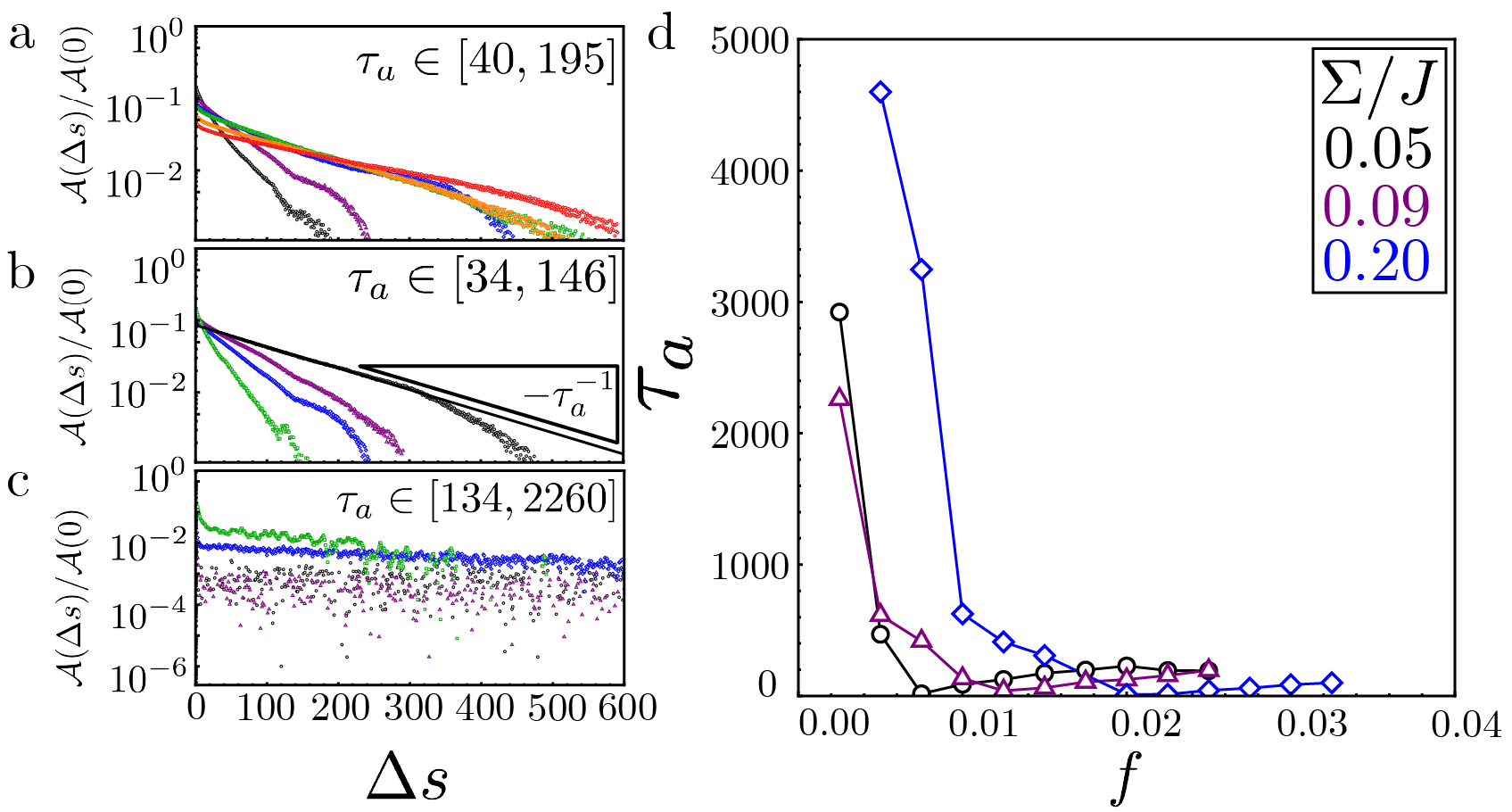}
\caption{\textbf{Auto-correlation functions} The auto-correlation function for the parameter set $\Sigma/J=0.09$, $\beta J=40.0$ and $L=800$ is computed for the sequences of \textbf{a} increasing strong frustration (with $\Phi=0.16$): $f=0.0125,0.015,0.0175,0.02,0.0225,0.025$ (black, purple, blue, green, orange and red, respectively), \textbf{b} increasing concentration ($f=0.015$): $\Phi=0.08,0.12,0.16, 0.2$ (black, purple, blue, green) and \textbf{c} increasing weak frustration ($\Phi=0.16$): $f=0.0025,0.005,0.0075, 0.01$ (black, purple, blue, green). Auto-correlation times $\tau_a$ are obtained by fitting $\ln {\cal A}$ to the intermediate time scale linear regime. \textbf{d} The auto-correlation times measured from Figure \ref{fig: autocorr}a-c plotted against frustration. From this we can see that, as frustration is lowered from strong to weak (i.e. where the system undergoes a condensation transition) the auto-correlation times grow large due to diverging time scales for macroscopic cluster formation and deformation.} \label{fig: autocorr}
\end{figure*}

\noindent {\bf Identification and characterization of aggregates} We identify {\it aggregates of subunits} as clusters of nearest neighbor occupied bonds in simulation snapshots using the Hoshen-Kopelman algorithm \cite{PhysRevB.14.3438}.  Referring to a particular subunit cluster (i.e. aggregate)  $c$,  $c(p)$ refers to the aggregate of subunit $p$, and $n_c$ is the number of subunits in $c$.  Given this definition, the mass distribution (for a given simulation snapshot) is simply
\begin{equation}
\phi_n = \frac{n}{A} \sum_c \delta_{n,n_c} .
\end{equation}
Equilibrium distributions for $\phi_n$ are generated by averaging mass-distributions of statistically uncorrelated snapshots, collected at intervals larger than the aggregation auto-correlation time.  For the simulations shown in Fig. \ref{fig: aggregation}b,  we sampled $\phi_n$ every 100 sweeps.  Given this averaged distributions, we characterize the mean aggregate size as 
\begin{equation}
\langle n \rangle_{\phi} = \sum_n n \phi_n /\Phi,
\end{equation}
as well as the aggregation fraction,
\begin{equation}
{\cal F} = \sum_{n> n_{\rm disp}} \phi_n/\Phi ,
\end{equation}
where $n_{\rm disp}$ is a cut-off to separate the disperse aggregates (i.e. monomers, dimers, trimers $\ldots$) from aggregates.  In Fig. \ref{fig: aggregation}b, this cut-off between disperse and aggregated states of subunits is characterized by an apparent minimum in $\phi_n$. For all simulations we set $n_{\rm disp}=9$ as the cut-off between disperse clusters and aggregated states of subunits.  

We characterize the finite size of aggregates in terms of their {\it effective width}, $w_{\rm eff}(p)$, which derives from the local distance of an aggregated subunit $p$ to nearest free boundary.  Defining $b(p)$ as the {\it shortest graph distance} from subunit $p$ to a subunit on the boundary of an aggregate (i.e. subunit missing at least one neighbor) and $p_\perp$ the set of subunits with nearest neighbor bonds in both $x$ {\it and} $y$ directions the effective width is
\begin{equation}
w_{\rm eff} (p) = \left\{ \begin{array}{ll} 4\big[ b(p)+\frac{1}{2} \big], & p \in p_\perp \\ \\ 1, & \ p \notin p_\perp \ \end{array} \right.
\end{equation}
Note the factor of $1/2$ for $p \in p_\perp$ places the boundary halfway between the occupied boundary site at the neighboring vacant site, and the factor of $4$ is introduced since the mean distance from the boundary of the rectangular strip of width $W$ is $W/4$.  The rule for $p \notin p_\perp$ is defined so that linear chains of occupied sites are properly counted as $w_{\rm eff} (p)=1$.  We compute the distribution of effective widths (Fig. \ref{fig: aggregation}c) simply from $\phi (W_{\rm eff}) = \frac{1}{A} \sum_p \delta_{W_{\rm eff},w_{\rm eff}(p)}$.

\noindent {\bf Intra-aggregate strain energy} We map and visualize the internal distributions (spin) elastic strain energy from finite-$T$ lattice simulations in terms of the deviation of rotation of neighbors from its preferred value.  For a given bond $ij$ the bond strain follows from expanding the spin dependent interactions in eq. (\ref{eq: Hamiltonian}) around its minimal energy value $\Delta\theta_{ij}=A_{ij}$, from this we define
\begin{equation}
\varepsilon_{ij} = \frac{1}{2}\lvert\Delta\theta_{ij}-A_{ij}\rvert^2 .
\end{equation}
For each given occupied site $i$, we compute the mean value of $\varepsilon_{ij}$ averaged over its occupied (i.e. bond) neighbor, to given the strain energy at site $i$, $\varepsilon (i)$.

At finite temperature and frustration, phase strain is generated both by frustration of the ground state, as well as thermal fluctuations (i.e. spin waves) around that state.  To better illustrate the underlying elastic gradients of intra-aggregate stress and compare them to ground state continuum theory predictions, as shown in the main text visualizations, we attempt to average our thermal fluctuations of spin as follows.  We computing intra-aggregate strains within a particular simulation snapshot, we temporarily turn off translational moves in the simulation, specifically cluster moves and SP translations and translations + rotations, and perform this spin-only MC regimen for $10^4$ sweeps, sampling the values of $\varepsilon (i)$ every $10^2$ sweeps.


\begin{figure*}
\centering
\includegraphics[width=0.85\textwidth]{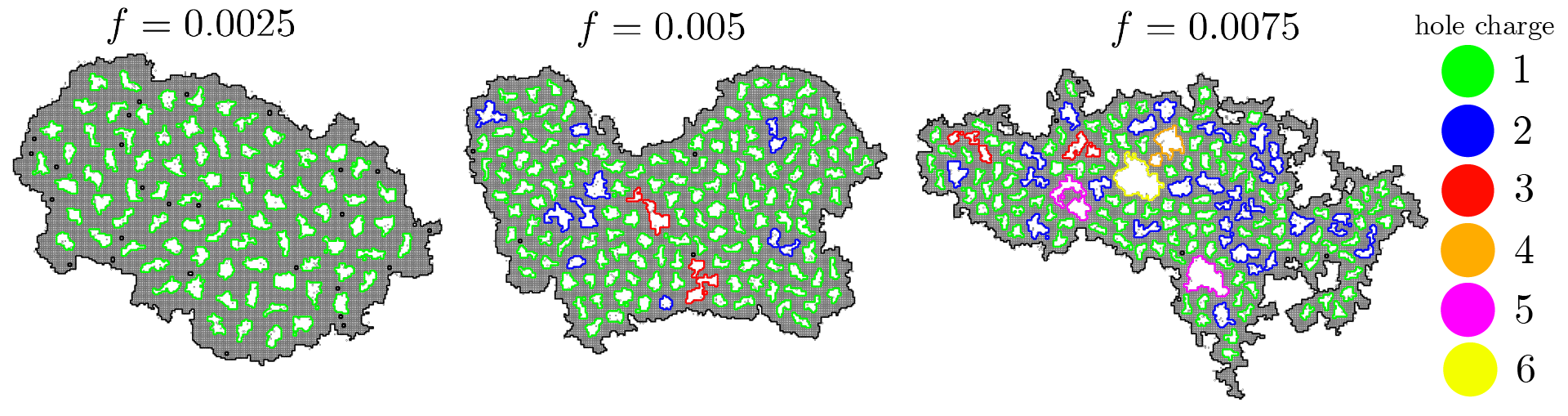}
\caption{\textbf{Topological charge of defect holes in vortex-sponge aggregates} Condensed aggregates (isolated) with interior hole boundaries colored by enclosed topological charge for frustration values of $f=0.0025,0.0050,\text{and }0.0075$ (with $\Sigma/J=0.09$ and $\beta J=40$). When $f\ll f_c$ we observe a uniform distribution of elementary ($q=1$) charge defects (ex. $f=0.0025$). For larger $f$ we observe higher $q=2,3,4\cdots$ charges (ex. $f=0.0050$), some of which may result of fusion of multiple $q=1$ holes. For $f\sim f_c$ (SI ex. $f=0.0075$) we observe that the size and charge distribution fluctuates, consistently larger holes have larger charge.}  \label{fig: Hole Defect}
\end{figure*}

\begin{figure*}
\centering
\includegraphics[width=0.85\textwidth]{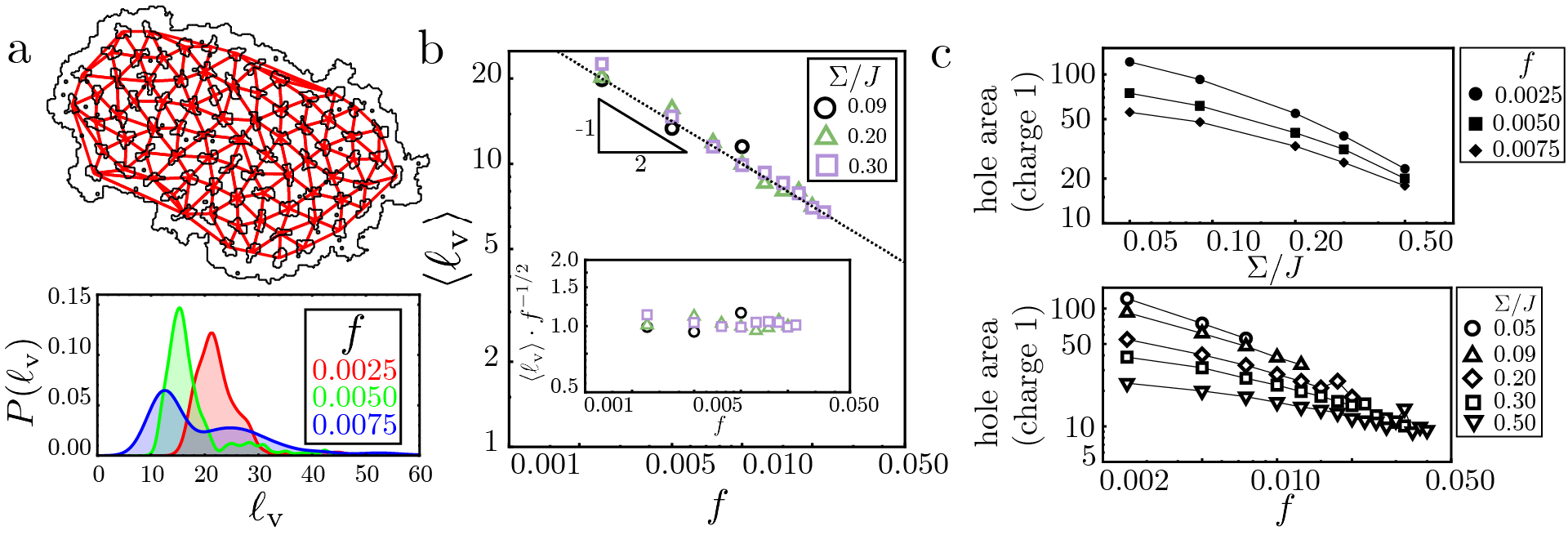}
\caption{\textbf{Characterization of defect hole separation and size} \textbf{a} Delaunay triangulation (top panel) of the $f=0.0025$ condensate where vertices only considered for centers of holes with $q=1$ charge. The bottom panel shows the distribution of inter-hole separation distances for the three example structures shown in Fig. \ref{fig: Hole Defect}. \textbf{b} Characteristic defect spacing (taken as mode of $P(\ell_{\rm v})$) as function of $f$ for three different values of $\Sigma/J$. The dotted line is the expected inter-vortex spacing for an Abrikosov defect lattice in the uniformly frustrated 2D XY model. The inset shows the characteristic defect spacing normalized by expected Abrikosov spacing $f^{-1/2}$, showing no evident variation with $\Sigma/J$. \textbf{c} Area of holes (with unit charge) for fixed frustration and increasing $\Sigma/J$ (top panel) and for fixed $\Sigma/J$ and increasing frustration (bottom panel).}
\label{fig: Hole sep}
\end{figure*}

\noindent {\bf Aggregation auto correlations} We measure the auto-correlation of aggregation to ensure equilibrium sampling.  Specifically, we compute the time scales (in MC simulation units) for aggregates to assemble and disassemble, to generate statistically independent populations of aggregates, following methods used for micellar simulations~\cite{von1997stochastic}.  
We define a tracer auto-correlation function:
\begin{equation}
\mathcal{A}(\Delta s)=\frac{\langle n_t(s+\Delta s) n_t(s)\rangle-\langle n_t(s)\rangle^2}{\langle n_t^2(s)\rangle-\langle n_t(s)\rangle^2} . 
\end{equation}
where $s$ and $\Delta s$ initial and intervals of time steps (or ``sweeps"), and $n_t$ is cluster size for a tracer particle, $p=t$ (i.e. $n_t = n_c(t)$ .  In practice we compute the auto-correlations by averaging $t$ over all particles.  Fig. \ref{fig: autocorr}a-c shows plots $\mathcal{A}(\Delta s)/\mathcal{A}(0)$, for a range of SLA simulation conditions, showing the decay to zero at long times, indicating that a given particle is likely to belong to a statistically uncorrelated aggregate at sufficiently long times (i.e. longer the the assembly/disassembly time).  While the observed relaxation is evidently not single-exponential, we fit $\mathcal{A}(\tau)\sim e^{-\Delta s/\tau_a}$ at intermediate times (longer than short-time single subunit exchange dynamics with free monomers)  and use $\tau_a$ as a measure of aggregation relaxation (i.e time scale for cluster formation/disassembly). Table S11 reports $\tau_a$ for a range of simulated parameters, varying $f$, $\Phi$, $\beta J$ and $\Sigma/J$.

\begin{figure*}
\centering
\includegraphics[width=0.85\textwidth]{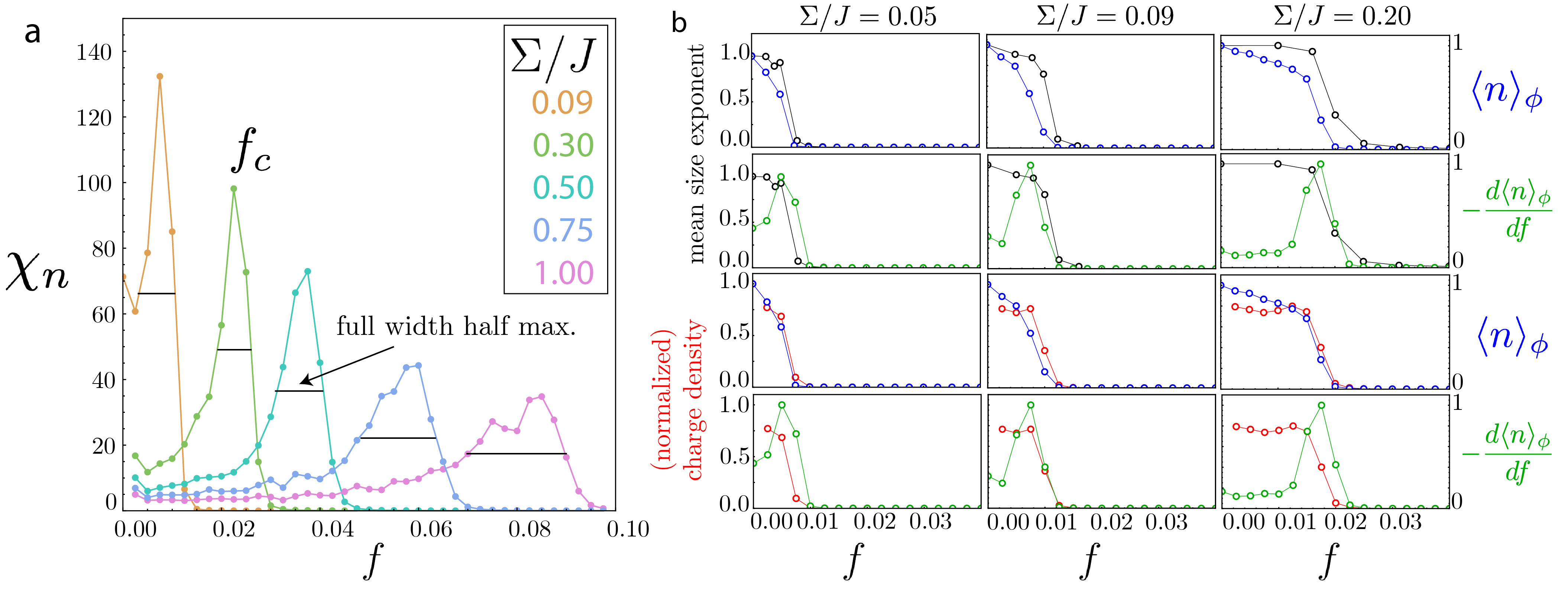}
\caption{{\bf Metrics for observing the condensation transition} \textbf{a} $\chi_n$ measured from simulations with $\Phi=0.1$, $\Sigma/J \in 0.09,0.3,0.5,0.75,1.0$ and fixed $\beta\Sigma=3.6$. The location of the peak of each distribution is used to identify the critical condensation frustration for the corresponding value of $\Sigma/J$ and the full width half maximum is used as a measure of the uncertainty. \textbf{b} Comparison of the different metrics used to observe the condensation transition for the three values of surface energy to elastic (spin) stiffness ($\Sigma/J=0.05,0.09,0.20$ and $\Phi=0.15$) used as examples Fig. \ref{fig: bulktodisperse}bc. The peak of the mean size susceptibility (green) lines up with the point where both the mean size exponent (black) and the normalized charge density (red) drop to zero; hence we use the location of this peak to define the critical value of frustration.}\label{fig: metric comparison}
\end{figure*}

\noindent {\bf Measuring aggregate charge} The topological charge, $q$, contained within a closed lattice loop of bonds, $\mathcal{C}$, can be measured through the relation\cite{wheatley1995flux}:
\begin{equation}\label{eq: HoleCharge}
\sum_{\mathcal{C}}\mod[\Delta\theta_{ij}-A_{ij}]=2\pi(q-f\mathcal{P})
\end{equation}
where $\mathcal{P}$ denotes the number of elementary plaquettes enclosed within the closed contour $\mathcal{C}$ and the gauge-invariant phase differences are defined in range $(-\pi,\pi]$. We construct orientated loops that enclose holes within clusters via a discretized Stokes law construction, in which clockwise oriented $+1$ ``currents'' are added to the bonds of all plaquettes with fully occupied vertices.  As bonds shared between two fully occupied plaquettes cancel, the superposition of integer charged loops leads to two sets of cycles within a cluster:  one (clockwise) ``macrocycle" enclosing the outer boundary of a cluster and a set of $n_{\rm hole}$ (counter-clockwise) cycles running around the edges of internal holes in the cluster.  

Using this set of oriented loops, we evaluate eq. (\ref{eq: HoleCharge}) and the number plaquettes enclosed in each loop (using $\sum_\mathcal{C} A_{ij}/(2 \pi f)=\mathcal{P}$), to compute the topological charge $q$ of each hole, as well the net charge of the entire aggregate.  A visual example of all of the holes\textemdash along with their measured topological charge\textemdash is shown in Fig.\ref{fig: Hole Defect} for several example aggregates.  An analysis of the center-to-center spacing $\ell_{\rm d}$ for variable frustration and cohesion to stiffness ratios is shown in Fig.\ref{fig: Hole sep}.

\noindent {\bf Aggregation-to-condensation transition} We performed finite-size scaling $\langle n \rangle_\phi \sim A^{\nu}$ analysis of mean cluster size $\langle n \rangle_\phi$ versus lattice size $A$ for only a limited set of parameters described in Fig. ~\ref{fig: bulktodisperse}b-c to characterize the bulk-to-disperse simulation.  For the broader range of parameters analyzed in Fig. ~\ref{fig: bulktodisperse}d, we performed a simpler analysis of this finite-size scaling, but at a fixed lattice size.  The approach follows from the fact that in the bulk state as $f \to 0$, the mean aggregate size scales with the largest aggregate and hence $\langle n( f \to 0) \rangle_\phi ~ A$, whereas at strong frustration, states are dispersed, and the mean aggregate size is independent of system size, and in generally much smaller than that bulk $f \to 0$ state.  Hence, as a function of increasing $f$, $\langle n( f ) \rangle_\phi/\langle n( 0 ) \rangle_\phi$ falls from ${\cal O}(1)$ in weak frustration to ${\cal O}(A^{-1})$ at the the transition point between bulk and dispersed states.  For large $A$ this transition is sufficiently sharp to measure a critical frustration, $f_{\rm c}$, from the generalized susceptibility, 
\begin{equation}
\chi_n = - \frac{1}{\langle n(0)\rangle_{\phi}}\frac{d \langle n(f)\rangle_{\phi}}{df} .
\end{equation}
In practice, and as shown in Fig. \ref{fig: metric comparison}a, we measure $f_{\rm c}$ from the peak value of a finite-difference approximation to $\chi_n$, and associate the uncertainty of its values with the variance.  Comparisons to finite-size scaling analysis for measurement of $f_{\rm c}$ in Fig. \ref{fig: metric comparison}b show quantitative agreement between the two measures within resolution limits allowed by discrete values of $f$ in the model.

\section{Mapping to models of physical, frustrated systems\label{mapping_appendix}}
\renewcommand{\thesubsection}{B.\arabic{subsection}}


Here, we summarize the mapping between our generic continuum model of geometrically frustrated assembly (eq. \ref{eq: continuum}) and two different models of distinct physical classes of frustrated systems: (i) liquid-crystalline (LC) assemblies on non-Euclidean surfaces; (ii) bend-nematic assemblies on planar substrates.  In both cases, the model parameters $\Sigma$ and $\epsilon_{\rm bulk}$ describe the (bare) free energy of boundaries and the bulk of aggregates, respectively. In general, both of these increase in magnitude with {\it cohesive interactions} between subunits.  More generally, the energetics of eq. (\ref{eq: continuum}) are applicable to any case where there is phase separation into a dense/ordered phase and a dilute/disordered phase, as is characteristic e.g. for lyotropic liquid crystalline states, in which $\Sigma$ parameterizes the line tension between dense and dilutes states.  The distinct mechanisms of frustration then enter in the form of the elastic energy and its effective mapping to the frustrated 2D XY model, described as follows.

\noindent {\bf LC assemblies on non-Euclidean surfaces} The elasticity theory of liquid-crystalline order embedded on surfaces of non-zero Gaussian curvature was derived by Nelson and Peliti~\cite{nelson1987fluctuations} and studied in the context of ``bulk'' phases that fully cover curved 2D manifolds by numerous others, including ~\cite{prost_lubensky, bowick, vitelli2006nematic}.  Here, following ref. ~\cite{nelson1987fluctuations} we employ the {\it Monge gauge}, which considers a surface whose shape is described by its height above the x-y plane, $h(\xv)$, i.e. the position is given by ${\bf R} (\xv) = x\hat{x} + y\hat{y}+ h(\xv)\hat{z}$ \cite{kamien2002geometry}.  In LC domains, order is described by a director $\nv (\xv)$ describing orientation in the tangent plane of the surface,
\begin{equation}
    \nv (\xv) = \cos \theta (\xv) ~ \ev_1 (\xv) + \sin \theta (\xv) ~ \ev_2 (\xv) 
    \label{eq: ndef}
\end{equation}
where $\ev_1 (\xv)$ and $\ev_2 (\xv) $ are defined to be orthonormal and perpendicular to the local surface normal $\ev_1 (\xv) \times \ev_2 (\xv) = \Nv (\xv)=(\hat{z} - \nabla_\perp h )/\sqrt{1+{|\grad_\perp h|^2}}.$   We consider the small slope approximation where $|\grad_\perp h| \ll 1$, which is consistent with domains having (small) finite size relative to curvature radii of the surface.  In this case, it is straightforward to show that the following set of vectors,
\begin{eqnarray}
\ev_1 &\simeq& \hat{x}\big(1 - |\partial_x h|^2/2\big) + \partial_x h~ \hat{z} - \partial_x h \partial_y h/2 ~\hat{y}  + \ldots \nonumber \\
\ev_2 &\simeq& \hat{y}\big(1 - |\partial_y h|^2/2\big)  + \partial_y h ~\hat{z} - \partial_x h \partial_y h/2 ~\hat{x} + \ldots \nonumber \\
\Nv &\simeq & \big(1 - |\grad_\perp h|^2/2\big)\hat{z}  - \partial_x h ~\hat{x} - \partial_y h ~\hat{y} + \ldots 
\end{eqnarray}
satisfy orthonormality and unit length up to ${\cal O} (|\grad_\perp h|^3)$.  The most simple and widely used generalization of the Frank elastic energy of LC order on 2D surfaces only considers {\it intrinsic gradients} of orientation, that is only elastic costs due to changes of the director in the tangent plane of the surface itself.  Hence, these intrinsic gradients are measured in terms of surface derivatives of the form,
\begin{eqnarray}
    \partial'_i \nv&=& \partial_i \nv - \Nv (\Nv \cdot \partial_i \nv) \nonumber \\
      &=& (\nv \times \Nv ) \big[ \partial_i \theta - A_i \big] ,
      \label{eq: ngrad}
\end{eqnarray}
where 
\begin{equation}
A_i(\xv) \equiv - \ev_2 (\xv) \cdot \big[ \partial_i \ev_1 (\xv) \big] = \ev_1 (\xv) \cdot \big[ \partial_i \ev_2 (\xv) \big] ,
\label{eq: spinconn}
\end{equation}
This {\it covariant derivative} of $\nv$ results directly from projection of the derivatives of eq. (\ref{eq: ndef}) into the tangent plane.  The field $A_i(\xv)$ is the so-called ``spin connection'' that accounts for the differences in local coordinates when computing director gradients.  It can be further shown that this covariant derivative of $\nv$, $\partial_i \theta - A_i$, is invariant under arbitrary changes of the local coordinate frame.  Specifically, for a given director field, arbitrary rotations of the frame vectors $\ev_1 (\xv)$, $\ev_2 (\xv)$ around the normal by $\delta \theta(\xv)$ transform the director angle as $\theta(\xv) \to \theta(\xv)+ \delta \theta (\xv)$ while the spin connection transforms as $A_i(\xv) \to A_i(\xv) + \partial_i \delta \theta(\xv) $, leaving the covariant derivative unchanged.  The simplest form of the Frank free energy assumes a single elastic constant $K$ for in-plane splay and bend, which form
\begin{equation}
F_{LC} = \frac{K}{2} \int d^2 \xv \sqrt{g} g^{ij} (\partial_i \theta - A_i )(\partial_j \theta - A_j ),
\label{eq: LC}
\end{equation}
where $g^{ij}$ is the inverse of the metric $g_{ij}= (\partial_i {\bf R} ) \cdot (\partial_j {\bf R} )$ and $g$ is its determinant.  Note that this form of intrinsic Frank elastic gradient free energy neglects {\it extrinsic} couplings between the director orientation and the curvature tensor, which we return to briefly in the context of self-assembled membranes.  

Following Nelson and Peliti~\cite{nelson1987fluctuations} we consider the lowest order couplings to surface shape in the small slope approximation, valid in the limit of large curvature radii, wherein eq. (\ref{eq: LC}) takes the simple form of the elastic energy in eq. (\ref{eq: continuum})
\begin{equation}
F_{LC} (|\grad_\perp h| \ll 1) \simeq \frac{K}{2} \int d^2 \xv|\grad_\perp \theta - {\bf A}|^2 ,
\end{equation}
where the spin-stiffness is the Frank elastic constant, $J \to K$, and the gauge field is the spin-connection, which is a 2D vector field in the small-slope limit
\begin{equation}
A_i(\xv)  \simeq \frac{1}{2} \big( \partial_x h \partial_i \partial_y h -\partial_y h \partial_i \partial_x h \big) .
\end{equation}
While ${\bf A} (\xv)$ itself depends on the coordinate choice (i.e. local rotations of the $\ev_1 (\xv)$, $\ev_2 (\xv)$ frame), its 2D curl depends only on the surface shape,
\begin{equation}
\grad_\perp \times {\bf A} \simeq {\rm det} ( \kappa_{ij} ) = K_G , 
\end{equation}
where $\kappa_{ij} \simeq \partial_i \partial_j h(\xv)$ is the curvature tensor in the small-slope approximation.  This establishes the connection between frustration in the 2D XY model and the Gaussian curvature, $f \to K_G$, which can be shown to hold more generally beyond the small-slope limit~\cite{kamien2002geometry}.  This shows that the present case of fixed and small $f$ in the lattice model corresponds to assembly of LC ordered domains on surfaces of \textit{fixed and constant Gaussian curvature} (e.g. spheres), with frustration decreasing with increasing curvature radii.  

Note that different symmetries of LC order would correspond to topological defects (disclinations) of different fundamental charge.  Consider the case where the LC order parameter is invariant under $\theta \to \theta + 2 \pi /p$:  $p=1$ (polar/tilted); $p=2$ (nematic); $p=4$ (tetratic); and $p=6$ (hexatic).   These variants, and the spectrum of disclinations they support, are captured by generalization of phase dependent interactions in the lattice model of the form $-J \cos \big[ p ( \Delta \theta_{ij} - A_{ij} ) \big]$.

This model can also be applied to the assembly of subunits into LC membranes that favor a non-Euclidean shape.  In addition to the intrinsic Frank elastic terms of the form of eq. (\ref{eq: LC}) such systems are modeled by additional terms that couple to {\it extrinsic geometry}.  These elastic terms can be derived from couplings between the director and the curvature tensor (e.g. $n_i \kappa_{ij} n_j$) consistent with the symmetries of LC order~\cite{Helfrich1988}.  Models of this form have been deployed, for example, to study the hyperbolic shapes of tilt bilayer membrane ribbons in a variety of LC materials~\cite{oda1999tuning, Selinger1996, selinger2001theory}.  For example, Ghafouri and Bruinsma have shown that, in chiral membranes, such extrinsic terms can be modeled (in the small-slope limit) as an effective bending energy~\cite{GhafBruin}
\begin{equation}
    F_{ext}\simeq \frac{B}{2} \int d^2 \xv \big( \kappa_{ij} - \kappa^{(0)}_{ij} \big)^2 ,
    \label{eq: ext}
\end{equation}
where $B$ is a bending modulus and
\begin{equation}
\kappa^{(0)}_{ij} = \left( \begin{array}{cc} 0 & \kappa_0 \\ \kappa_0 & 0 \end{array} \right)
\end{equation}
is the preferred curvature tensor, with $\kappa_0$ a preferred off-diagonal curvature.  More generally, it can be shown that rotations of the molecular order can also be described in this form, but with rotations of preferred principle directions relative to coordinate frame~\cite{armon2014shape}.  For sufficiently large bending modulus, the extrinsic energy of the form of eq. (\ref{eq: ext}) will lock the curvature into a fixed value, i.e. $\kappa_{ij} = \kappa_{ij}^{(0)}$, such that frustration is {\it quenched} to a fixed, non-zero, $K_G = {\rm det} \big(\kappa_{ij}^{(0)}\big)=-\kappa_0^2$.  

In practice, frustration in such terms cannot be considered fixed, and the extrinsic elasticity of the form of eq. (\ref{eq: ext}) describes ``soft elasticity'' of the frustrating gauge field.  As a result, the effective frustration will adjust itself with the size and shape of the self-organized domain, which can be seen by considering two limits.  First, when the shaped is locked into the preferred one, the elastic energy density of a finite domain grows as $\sim K \kappa_0^4 W^2$.  Secondly, when the shape flattens to avoid the cost of frustration at the expense of bending energy density $\sim B \kappa_0^2$.  These two energy scales crossover at characteristic scale $W_{\rm flat} \sim \kappa_0^{-1} \sqrt{K/B}$, suggesting that for sufficiently large domains ($W \gg W_{\rm flat}$) the assembly can avoid the super-extensive costs of frustration {\it without defects} by elastic defrustation. This effect is not captured by the fixed-frustration regime that is the central focus of the present study.  Nevertheless, as the frustration relaxation is size dependent, we can expect it to have minimal impact on the regime of narrow, self-limiting domains (i.e. the regime of $f \gg f_c$).

\noindent {\bf Bend-nematic assembly on planar surfaces} Following Efrati and coworkers, we show that cohesive assembly of ``bend-nematic'' (BN) phase also exhibits frustration captured by a variant of the elastic energy in eq. (\ref{eq: continuum}).  A BN phase is a variant of nematic order where the director has a favorable bend. Here we consider a 2D scenario, appropriate, for example, for assemblies of banana shaped colloidal particles\cite{fernandez2021hierarchical}, where the 2D director $\nv (\xv)$ is embedded in-plane (itself a variant of the 3D case with a plethora of bend- or splay-nematic phases \cite{fernandez2020shaping,kotni2022,SelingerRevModPhys.90.045004}):
\begin{equation}
\nv(\xv) = \cos \theta(\xv) ~ \hat{x} + \sin \theta(\xv) ~\hat{y} .   
\end{equation}
A 2D domain of planar bend-nematic is described by the Frank free energy
\begin{equation}
F_{\rm BN} = \frac{1}{2} \int d^2 \xv \Big\{ K_1 (\grad_\perp \cdot \nv )^2 + K_3 (\grad_\perp \times \nv - b_0)^2 \Big\}, 
\label{eq: BN}
\end{equation}
where $K_1$ and $K_3$ are the Frank elastic constants that couple to the {\it splay} and {\it bend} of the director field and $b_0$ is the {\it preferred bend}, which is assumed to be uniform, an intrinsic property set by e.g. the shape of bent-core molecule or banana-shaped particle.  For this elastic energy, the texture favors everywhere to be splay-free and uniformly-bending (i.e. $\grad_\perp \cdot \nv =0$ and $\grad_\perp \times \nv = b_0$), a scenario which is easily generalized to arbitrary combinations of preferred splay and/or bend.  Remarkably, even on a Euclidean surface, such an assembly will experience geometric frustration for any non-zero preferred splay {\it or} bend.  We demonstrate this general result by mapping the bend-nematic elastic free energy eq. (\ref{eq: BN}) to the frustrated 2D XY model for the simplest single-elastic constant model $K=K_1=K_3$.  Defining a co-director $\mv \equiv \hat{z} \times \nv$ perpendicular to $\nv$, the local bend $b=\grad_\perp \times \nv= (\grad_\perp \theta) \cdot \nv$ and splay $s=\grad_\perp \cdot \nv=(\grad_\perp \theta) \cdot \mv$ characterize the gradient of the director angle,
\begin{equation}
\grad_\perp \theta = b(\xv) ~ \nv(\xv)+ s(\xv) ~ \mv(\xv) .
\end{equation}
Using this form and the single Frank constant assumption, eq. (\ref{eq: BN}) takes the form
\begin{equation}
F_{\rm BN} = \frac{K}{2} \int d^2 \xv \big| \grad_\perp \theta - b_0 \nv|^2 .
\end{equation}
Comparison to eq. (\ref{eq: continuum}) shows the planar bend-nematics map to the frustrated 2D XY model under $J\to K$ and with a gauge field,
\begin{equation}
    {\bf A}_{\rm BN} = b_0 \nv(\xv) .
\end{equation}
The effective frustration in the model is characterized by the 2D curl
\begin{equation}
\grad_\perp \times {\bf A}_{\rm BN} = b_0 b(\xv) .
\end{equation}
From this we find that, unlike the case of 2D nematics on non-Euclidean surfaces where frustration is controlled by additional shape degrees of freedom, for the planar bend-nematic phases, frustration is {\it self-generated} by the intrinsic preference for bend, which favors $b(\xv) = b_0$.  Hence, in the regime where the preference for the ground-state bend is sufficiently strong, we expect $f \to b_0^2$, which is consistent with the results of Efrati and Niv who showed that the sum of the squares of bend and splay act as the {\it geometric charge} for frustration that is analogous to the role played by $K_G$ surfaces for 2D nematics~\cite{NivEfrati}.  In this case, we see that the gauge field itself is fluctuating, like the case of the nematic membrane, and as such, the strength of frustration will vary with domain shape and size.  This is consistent with results of Meiri and Efrati \cite{meiri2022XY} who found that, in absence of topological defects, optimal bend of BN domains adopt the preferred value $b(\xv) = b_0$ in the limit of narrow domains, which the texture unbends to $b(\xv) < b_0$ to relax the super-extensive costs of frustration as domain sizes grow.  This represents an alternative, elastic mode of frustration escape that will compete for thermodynamics stability in the limit of low frustration, i.e. low $b_0$.    

\section{Continuum aggregate model of finite width aggregates \label{continuum_appendix}}
\renewcommand{\thesubsection}{C.\arabic{subsection}}

\noindent {\bf Energetics of rectangular domains} As a model of self-limiting aggregates of the lattice model, we consider strip-like rectangular domains of occupied subunits, with a minimal phase energy described by ground states of the continuum energy in eq. \ref{eq: continuum}.  Specifically, we consider domain shapes, ${\cal D}$ that occupy a region $x\in[0,X]$ and $y\in[0,Y]$ and consider Euler-Lagrange equations for $\theta({\bf x})$,
\begin{equation}
\nabla_\perp ^2\theta(x,y)=\nabla_\perp \cdot\mathbf{A}(x,y) = 0 ,
\end{equation}
with the boundary condition:
\begin{equation}
\hat{n}\cdot\nabla_\perp\theta(x,y)\big\rvert_{ \partial \mathcal{D}}=\hat{n}\cdot\mathbf{A}(x,y) ,
\end{equation}
where we have used the divergence free gauge, and $\hat{n}$ is normal to the domain boundary $\partial \mathcal{D}$.

Solving this equation on rectangular domains of dimension $X\times Y$ and including the bulk and boundary energetics (see Supplementary Sec. 1), we have the energy density of rectangular aggregates:
\begin{equation}
\label{eq: ED}
\epsilon(X,Y)=Jf^2\Omega\bigg(\frac{X'}{Y'}\bigg) ~ (X')^2-2\Sigma+\frac{X+Y}{XY} \Sigma
\end{equation}
where $\Omega(\alpha )$ is a dimensionless factor dependent on the aspect ratio of our aggregates
\begin{equation}
\Omega (\alpha)=\frac{\pi^2}{6}-\frac{32}{\pi^3} \alpha \sum_{\text{n odd}}\frac{1}{n^5}\tanh\big(\frac{n\pi }{2\alpha}\big).
\end{equation}
Notably, the elastic terms are function of $X'=X-1$, $Y'=X-1$ to account for the fact that the elastic energy is defined on the network of {\it bond} between sites, consistent with the intuitive result that phase strain, and elastic energy must vanish for $X,Y \to 1$.  In the regime where domain sizes are large compared to lattice dimensions, gross features of the energy density landscape are only weakly dependent on the microscopic lattice cutoff, and hence, we can make the approximation $(X',Y') \to (X,Y)$ in eq. (\ref{eq: ED}) above.  In Supplemental Sec. 2, we describe the quantitative corrections of the finite lattice cutoff.  In this limit, it is straightforward to show that the energy can be rescaled in a dimensionless form by measuring lengths in units of $\ell_{\rm d}$, $(\bar{X},\bar{Y})=(X/\ell_{\rm d},Y/\ell_{\rm d})$, yielding
\begin{equation}
\label{eq: EDscaled}
\frac{\epsilon(\bar{X},\bar{Y})-\epsilon_{\rm bulk}}{\Sigma/\ell_{\rm d}}=\Omega \big( \bar{X}/\bar{Y}\big)\bar{X}^2 +\frac{1}{\bar{X}}+\frac{1}{\bar{Y}} .
\end{equation}
It is straightforward to analyze the minimal energy shapes via the minimization of the substitution $\bar{X}=\sqrt{\alpha \bar{n}}, \bar{Y}=\sqrt{\alpha^{-1} \bar{n}}$ into eq. (\ref{eq: EDscaled}) and minimizing over $\alpha$ at fixed aggregate size $\bar{n}=\bar{X} \bar{Y}$, leading to the ``pitchfork'' saddle path states highlighted in Fig. \ref{fig: aggregation}a, with anisotropic states becoming favored for $n > n_{\rm c}$.

In the limit of large $\bar{n}$ optimal shapes approach $\alpha \to 0$, with ribbons approaching a constant finite width $W_* (n \to \infty)= (3/\pi^2)^{1/3} \ell_{\rm d}$, and obtaining limiting form of the energy density for {\it 1D strips}, 
\begin{equation}
\label{eq: ep1D}
\epsilon_{\rm 1D} (n) = \lim_{n \gg n_c} \epsilon(W_*,n/W_*) \simeq \epsilon_{\infty} + \Delta / n
\end{equation}
where $\epsilon_{\infty} = - 2 \Sigma + \frac{\Sigma (3 \pi)^{2/3}}{2} \Big( \frac{f^2}{\Sigma/J} \Big)^{1/3}$ is the limiting per-subunit energy in bulk of strips, and $\Delta = \Sigma/W_*$ effectively parameterizes the cost of finite-length strip ends.

\noindent {\bf Ideal aggregation of self-limiting domains} We use ideal aggregation theory to predict the aggregate distributions from the continuum theory model of the per subunit aggregation energy.  In particular, we consider the {\it minimal energy density} aggregates $\epsilon_{\rm m} (n) \equiv {\rm min} \big[ \epsilon (\sqrt{\alpha n},\sqrt{\alpha^{-1} n}) \big]$.  The equilibrium distributions in this regime are given by the law of mass action~\cite{HaganGrason}
\begin{equation}
\label{eq: massaction}
\phi_n (\mu) = n \big( e^{\beta [\mu - \epsilon_{\rm m} (n)]} \big)^n ,
\end{equation}
where $\mu = \ln \phi_1$ is the chemical potential of subunits, which is determined from the equation of state
\begin{equation}
\Phi = \sum_{n=1}^{\infty}\phi_n (\mu) .
\end{equation}
Note, this sum accounts for the two degenerate branches of anisotropic ``worm-like" aggregate states.  For the aggregation distributions shown in Fig.~\ref{fig: aggregation}b, we evaluate this sum by numerical integration along the contour of minimal energy aggregates, including the explicit effect of lattice cut-off (see SI sec. 2 for details). 

As aggregate populations are ultimately dominated by quasi-1D aggregates of finite width and, as shown in Fig.~\ref{fig: aggregation}b depleted of aggregate states intermediate to dispersed monomers to aggregates of size $n \gtrsim n_{\rm c}$ , a good approximation of the aggregation transition is given by a simplified model that considers only two populations of subunits:  free monomers $\phi_1$; and 1D aggregates for $n\geq n_{\rm c}$ with the energy $\epsilon_{\rm 1D} (n)$ given in eq. (\ref{eq: ep1D}).  Inserting this into the equation of state, we have,
\begin{equation}
\Phi(\phi_1) = \phi_1+ \Phi_{\rm agg}(\phi_1)
\end{equation}
where
\begin{multline}
\label{eq: agg}
    \Phi_{\rm agg}(\phi_1)=2 \int_{ n_{\rm c} }^{\infty} dn ~n  z(\phi_1)^n e^{-\beta \Delta}  \\ =2 e^{-\beta \Delta} z(\phi_1)^{n_{\rm c}} \frac{n_{\rm c} \ln z(\phi_1)^{-1}+1 }{[\ln z(\phi_1)^{-1}]^2}
\end{multline}
and $z (\phi_1)= \phi_1 \exp(-\beta \epsilon_{\infty}) $.  It is straightforward to see that there are two regimes with increasing $\phi_1$.  For low $\phi_1$ (dilute), $z \ll 1$ and $\phi_1\gg \Phi_{\rm agg}$ corresponding to the disperse monomer state.  When $\phi_1$ approaches
\begin{equation}
\phi_* = e^{\beta \epsilon_{\infty}}
\end{equation}
$z \to 1$, and it is clear to see that the population of aggregates diverges corresponding to the saturated regime $ \Phi_{\rm agg}\gg \phi_1$.  Noting that, at the crossover between monomer and aggregate dominated states, $\Phi_* \approx \phi_*$, we have a simple expectation for the critical aggregation concentration,
\begin{equation}
\label{eq: CAC}
(\beta \Sigma)^{-1} \ln \Phi_* \approx -2 + \frac{(3 \pi^2)^{1/3}}{2} \Big(\frac{f}{\sqrt{\Sigma/J}} \Big)^{2/3} ,
\end{equation}
Which defines the two key axes of the aggregation behavior in Fig. \ref{fig: aggregation}d.  Contours of constant aggregate fraction in this plot are generated by the 1D aggregation model, which we show in Fig. S5 to capture the key dependence of the full (variable aggregate aspect ratio) continuum model.


\bibliography{sn-bibliography}

\end{document}